**The Dynamics of Attention across Automated and Manual Driving Modes:**
**A Driving Simulation Study**

AUTHORS

Yuan Cai[1], Mustafa Demir[2], Farzan Sasangohar[2], Mohsen Zare*[1]

1- *UTBM, ELLIADD-ERCOS, Belfort Cedex, France*
2- *Department of Industrial and Systems Engineering, Texas A&M University, College Station, TX, USA*

**\*Corresponding Author:**
Name: Mohsen Zare
Department: *ELLIADD-ERCOS*
University: *UTBM*
Address: UTBM, 90100, Belfort, Cedexe
Email: Mohsen.zare@utbm.fr

**ACKNOWLEDGEMENTS**

This study was conducted within the framework of the European Union's Horizon 2020 research and innovation program and the PAsCAL project (Grant No. 815098). We thank Maxime Larique, Nicolas Bert, and Florian Girardot for their assistance with data collection.

**ABSTRACT**

**Objective:** This study aims to explore the dynamics of driver attention to various zones, including road, central mirror, embedded Human-Machine Interface (HMI), and speedometer, across different driving modes in AVs.

**Background:** The integration of autonomous vehicles (AVs) into transportation systems has introduced critical safety concerns, particularly regarding driver re-engagement during mode transitions. Past accidents underscore the risks when drivers overly rely on automation and highlight the need to understand dynamic attention allocation to support safety during autonomous driving.

**Method:** A high-fidelity driving simulation was conducted. Eye-tracking technology was utilized to measure Fixation duration, fixation count, and time to first fixation across distinct driving modes (automated, manual, and transition modes) were used to assess how drivers allocated attention to various areas of interest (AOIs).

**Results:** Findings show that drivers' attention varies significantly across driving modes. In manual mode, attention consistently focuses on the road, while in automated mode, prolonged fixation on embedded HMI was observed. During the Handover and takeover phases, attention shifts dynamically between environmental and technological elements.

**Conclusion:** The study reveals that driver attention allocation is mode-dependent. These findings inform the design of adaptive HMIs in AVs that align with drivers' attention patterns. By presenting relevant information according to the driving context, such systems can enhance driver-vehicle interaction, support effective transitions, and improve overall safety.

**Application:** Systematic analysis of visual attention dynamics across driving modes is gaining prominence, as it informs adaptive HMI designs and driver readiness interventions. The GLMM findings can be directly applied to the design of adaptive HMIs or driver training programs to enhance attention and improve safety.

**Keywords:** Attention, Autonomous driving, Eye Movement, Human-Automation Interaction, Human-Machine Interface.

**INTRODUCTION**

In recent years, the rise of automated and connected vehicles has promised significant economic and societal transformations. Projections estimate that these technologies could contribute up to $1.2 trillion annually in the United States (U.S.) alone(Clements & Kockelman, 2017), This technology will offer new levels of independence to individuals unable to drive, such as the elderly and disabled people, empowering them to participate more in society (Huff et al., 2019; Mesquita et al., 2023). Furthermore, autonomous systems hold great potential for improving traffic law compliance, with some studies predicting a significant increase in adherence to regulations, from 13% to over 80% (X. Ma et al., 2023; Prakken, 2017). By reducing human error and improving traffic safety, autonomous vehicles could drastically reduce collisions (Wang et al., 2020).

Despite the promise, the introduction of autonomous technology has also raised concerns, due to occasional fatal accidents. One of the earliest examples of this is the fatal 2018 crash involving an Uber self-driving car, underscore the dangers when drivers fail to stay attentive during autonomous operation in Arizona, U.S.(Brainard, 2018). Other ones, including the 2016 Tesla Model S crash and the 2018 Tesla Model X highway barrier collision, highlight the risks of driver over-reliance on automation. In these cases, despite receiving warnings, drivers failed to remain attentive, and the vehicle systems were unable to respond effectively (Banks et al., 2018; Canellas & Haga, 2020). These accidents reveal a crucial challenge: maintaining appropriate driver attention during autonomous driving. In highly automated vehicles, designed to function without continuous driver supervision, drivers often shift their focus to non-driving tasks after handing over control. Studies have shown that even with less advanced systems, like Advanced Driver Assistance Systems (ADAS), drivers tend to pay less attention to the road (Pipkorn et al., 2024). Yet, when autonomous systems encounter situations beyond their capabilities, drivers must be ready to take over control. The ability to respond promptly to Take Over Requests (TORs) depends on drivers maintaining adequate attention throughout the drive, a factor that varies greatly among individuals.

**Attention in Driving**

Visual attention during autonomous driving plays a critical role in safely navigating unexpected events. Visual attention, defined as the cognitive process of selecting relevant information from complex visual scenes while filtering out irrelevant details (McMains & Kastner, 2009), has been commonly measured through eye-tracking technology. Variables such as fixation duration, fixation count, and the time to first fixation provide insights into drivers' allocation of visual/spatial attentional resources and gaze behavior. Research has further explored how visual attention shifts during the transition between autonomous and manual driving. Pipkorn et al. (2024) found significant differences in 'drivers' gaze behavior across different Areas of Interest (AOIs), such as the road and the 'vehicle's Human-Machine Interface (HMI) during manual and autonomous modes and during transitions. Drivers directed significantly less gaze toward the road (41.2%) during autonomous driving compared to manual driving (75.4%). When TORs occurred, attention temporarily shifted to the instrument, reaching 40% before decreasing to below 30% once manual control was reestablished.

Ma et al. (2020) demonstrated that cognitive workload and visual attention are in competition, with cognitive processes often taking precedence over visual attention during periods of high mental demand. Similarly,  showed that gaze behavior, particularly fixation count, is heavily influenced by the nature of Non-Driving-Related Tasks (NDRTs) performed prior to a TOR.

**Trust in Driving Automation**

Visual attention during autonomous driving varies among drivers, with some frequently looking away from the road while others maintain focus. This variation may be influenced by factors such as trust in automation (Hergeth et al., 2016). Drivers with higher levels of trust in automated vehicles are more likely to reduce their visual attention to the road and other monitoring efforts (Brown & Galster, 2004). Although trust is essential for the effective use of automation, excessive trust can lead to disengagement from critical driving tasks, impairing the ability to respond to emergency situations. Llaneras et al. (2013) showed that participants paid less attention to the road when assisted by multiple automation systems compared to simpler systems. To optimize the interaction between drivers and autonomous systems, a well-calibrated trust is necessary. As autonomous driving systems relieve drivers from the driving task, they can diminish drivers' situational awareness, reducing their ability to perceive, understand, and predict the traffic environment (Zihui et al., 2022). Studying the dynamics of drivers' attention can offer valuable insights into driving strategies, behaviors, and cognitive processes, which is crucial for improving the safety of driver-autonomous system interactions and preventing potential accidents.

Particularly, the interaction between visual attention and 'drivers' trust in automation warrants further investigation. Most previous studies have focused on participants engaging in NDRTs (Guo et al., 2024; Hecht et al., 2020; Liang et al., 2021, 2021; Peng & Iwaki, 2020; Pipkorn et al., 2024; Schartmüller et al., 2021), but this study uniquely examines visual attention in the absence of NDRTs, where participants must remain focused on driving tasks and respond to TORs even in automated mode.

**Research Questions and Hypotheses**

This study investigates two main research questions to provide new insights into the complex relationship between visual attention, trust, and driver behavior in the context of automated driving:

**(1)** *Does attention, as measured by eye-tracking, vary across Areas of Interest (AOI) of technological and environmental factors over time within the four driving modes, including automation, Handover, manual, and takeover?*

*Hypothesis 1:* It is hypothesized that attention will vary significantly across AOIs over time, depending on the driving mode. In manual driving mode, greater attention will be allocated to environmental factors due to the higher cognitive demands of active driving. In automated driving mode, more stable fixation durations will be observed on technological factors (e.g., Tablet, speedometer). During handover mode, attention will shift toward both technological and environmental factors as control transitions to the driver, while takeover mode will involve rapid attention shifts between these factors to regain control. Overall, manual and takeover modes are expected to show more dynamic attention shifts than automated and Handover modes.

**(2)** *How does the level of trust in driving impact attention allocation to environmental and technological factors across the four driving modes: automation, Handover, manual, and takeover?*

**Hypothesis 2:** It is hypothesized that as trust increases, participants will show reduced attention to monitoring-related AOIs (e.g., speedometer, central mirror), particularly in automation mode. At the same time, higher trust levels are expected to shift attention toward secondary tasks, such as tablet-related AOIs, resulting in longer fixation durations and more frequent fixations on these tasks. The relationship between trust and attention is expected to be nonlinear (Tenhundfeld et al., 2022), with higher trust levels leading to more pronounced shifts in attention.

## METHODS

### Participants

An a priori power analysis was conducted using G*Power3 (Faul et al., 2007) to test the means of two conditions across seven simulation events using an F-test with a medium effect size ($\eta^2_p = 0.06$; Cohen, 1988), and an alpha ($\propto$) of 0.05. The analysis determined that a sample of 26 participants was required to achieve a power of 0.95. To account for potential missing data, we initially recruited 40 randomly selected participants drawn from the largest Eastern France University and its surrounding community for the experiment. However, six participants could not complete the tasks, resulting in a final sample of 34 participants. The resulting demographics collected at the end of the study show that participants were generally younger adults ($M_{age}$= 33 and $SD_{age}$= 16; Male = 15 and Female = 19).

Participants were selected based on the following inclusion criteria: possession of a valid driving license, regular driving experience of at least 3,000 km in the past year, fluency in French, good general health, having normal or corrected-to-normal hearing and vision (e.g., colorblind participants were excluded from this study), no history of epilepsy, and an age range of 18 to 85 years, and be comfortable using a standard computer mouse and keyboard. They were compensated 10 Euro per hour for their time spent during the study, or roughly 30 Euro (€) total.

## Simulation Driving Testbed

The simulation environment comprises three key elements: (1) the simulator, (2) the PCs used to control the simulator, and (3) a tablet utilized to display the HMI to the user. The simulator features a panoramic projection screen and a modified Peugeot 308 vehicle, with its powertrain removed (Figure 1a). The vehicle has sensors, force feedback mechanisms, and electronic components to interface seamlessly with the simulation PC. The simulator provides participants with an immersive environment with a realistic driving experience. It features a cabin and front screen setup that simulates the interior and view of a car, enabling participants to interact with vehicle controls and the road environment in a way that closely mimics real-life driving (Figure 1b). The simulation software runs on a PC that is part of a cluster. Four other PCs are dedicated to computing virtual views and displaying them through three beamers composing a 180° front view on a large cylindrical screen facing the car and three monitors facing the mirrors for rear-viewing. The simulator is capable of simulating Level 3 autonomous driving, allowing transitions between manual and autonomous driving modes.

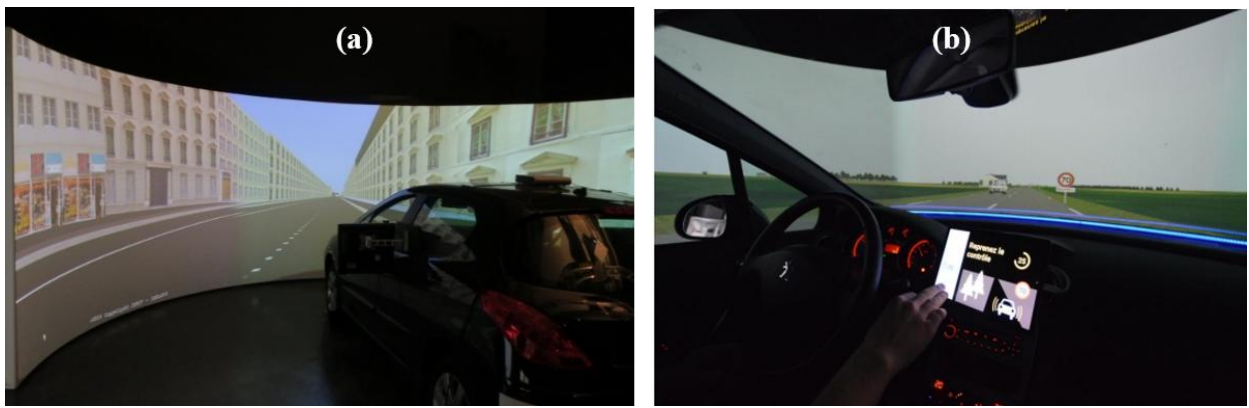

**Figure 1. (a)** Panoramic projection screen with a real Peugeot 308 vehicle; **(b)** Driver interacts with HMI tablet during simulation.

## Embedded HMI design

The 'simulator's multichannel HMI was developed using an Android tablet (a 10-inch Android tablet), with an Arduino Mega 2560 interface between the simulator, PC, and Tablet. The Arduino forwards events between these components and controls the dashboard LEDs based on tablet instructions. The tablet display was designed based on a benchmarking analysis of existing automotive HMIs, ensuring alignment with industry practices and traffic regulations. The HMI is intentionally simplistic, focusing on critical information to minimize distraction while emphasizing key autonomous driving events through dashboard LEDs, audio/voice messages, and touchscreen text notifications.

The HMI delivered a variety of signals including visual, light, auditory, and vocal cues to communicate with the driver about the autonomous driving activity. It also provided a control button for taking over or handing over driving responsibilities and informed the user about the 'vehicle's environmental perception. To simulate feedback regarding the 'vehicle's perception of its environment, the designed HMI displays the following information: the current road type (urban, rural, or highway), the current speed limit, potential upcoming events (such as traffic jams, obstacles (broken cars), or roadwork), the Connected Autonomous Vehicle (CAV)'s level of environmental awareness (excellent, medium, or poor), and the status of autonomous driving availability (Figure 2).

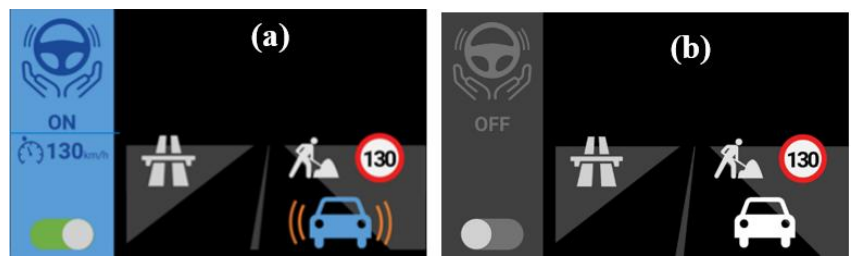

**Figure 2.** Screenshots of the touchscreen tablet interface under different conditions: **(a)** Feedback display during active automation, indicating the current legal speed limit (130 km/h), road type (highway), an upcoming event (road work), and a medium perception level; **(b)** Feedback display when autonomous driving is disabled.

**Simulation Scenario**

The simulation scenario assessed user interaction with an autonomous driving system under varying road conditions and traffic situations (Figure 3). The scenario begins with the subjects starting the engine and driving manually on a clear country road. After entering a suburban area, the HMI indicates that autonomous driving is available, which the subjects activate. The car then navigates through a rural area with increasing traffic, encounters roadworks where it automatically adjusts speed and trajectory, and exits the roadworks to resume the legal speed. At various points, the HMI instructs the subjects to take control, which they do by steering or pressing a pedal. The subjects then enter a highway, with the HMI reactivating autonomous driving as traffic conditions change, including a traffic jam where the car automatically adjusts speed. After navigating through a series of road conditions, including another roadworks area and a traffic jam, the HMI prompts the subjects to take control multiple times, with autonomous driving reactivated as needed. The scenario concludes with the car autonomously parking on the roadside and the subjects turning off the engine.

**Figure 3.** Simulation scenarios for assessment of autonomous driving interaction in varied road conditions.

**Experimental Design**

**Procedure**

Due to the COVID-19 pandemic, participants were required to register online and wear a facemask during their visit to the university facilities. Upon arrival, participants were briefed on hygiene protocols and the experimental procedure. They then completed a demographic questionnaire on personal information, driving experience, and attitudes toward autonomous vehicles. The researcher then assisted participants in wearing the necessary measurement devices, including Galvanic Skin Response, Novocar heart-beat measurement, and mobility wrist measurement.

Participants familiarized themselves with the driving simulator by adjusting the cockpit to their preferences and engaging in a 15-minute driving session, followed by a 5-minute rest period to record baseline physiological data. The experiment involved driving a Level 3 autonomous vehicle through various environments, with several Handover and takeover maneuvers prompted by different signals in the scenario. Following the experiment, sensors were removed, and participants engaged in a semi-structured interview discussing their experience, trust in the autonomous vehicle, and the effectiveness of the signals.

**Equipment and Apparatus**

To evaluate gaze behavior during driving, 'subjects' eye movements across various AOIs on the HMI tablet, road, and speedometer (Figure 4) were recorded using the Tobii Pro Glasses 2 eye-tracking system. This system consists of a wearable head unit resembling standard glasses and a recording unit that captures eye movement data during driving tasks. The Tobii Glasses Controller Software was used to calibrate the system, manage recordings, handle participant data, and define AOIs. The recording frequency, which determines the number of gaze points captured per second, was set at 50 Hz. When analyzing the eye-tracking data, we initially segmented the visual field into ten distinct AOIs to capture detailed gaze patterns, as illustrated in Figure 4. These comprised four AOIs within the windshield area, including landscape left and right, light band, and road; four within the tablet interface, one for the speedometer, and one for the central mirror. However, to simplify the data structure and to understand attention allocation at a more actionable, broader level, we decided to pool the AOIs within the windshield and the Tablet into two overarching categories: one encompassing all areas within the windshield and the other encompassing the entire tablet area. Focusing on these larger functional areas aligns better with our primary research objective of understanding how drivers allocate their visual attention between the primary driving environment (windshield) and secondary in-vehicle systems (Tablet). Furthermore, pooling the AOIs enhanced the statistical robustness of our findings by increasing the amount of data within each category.

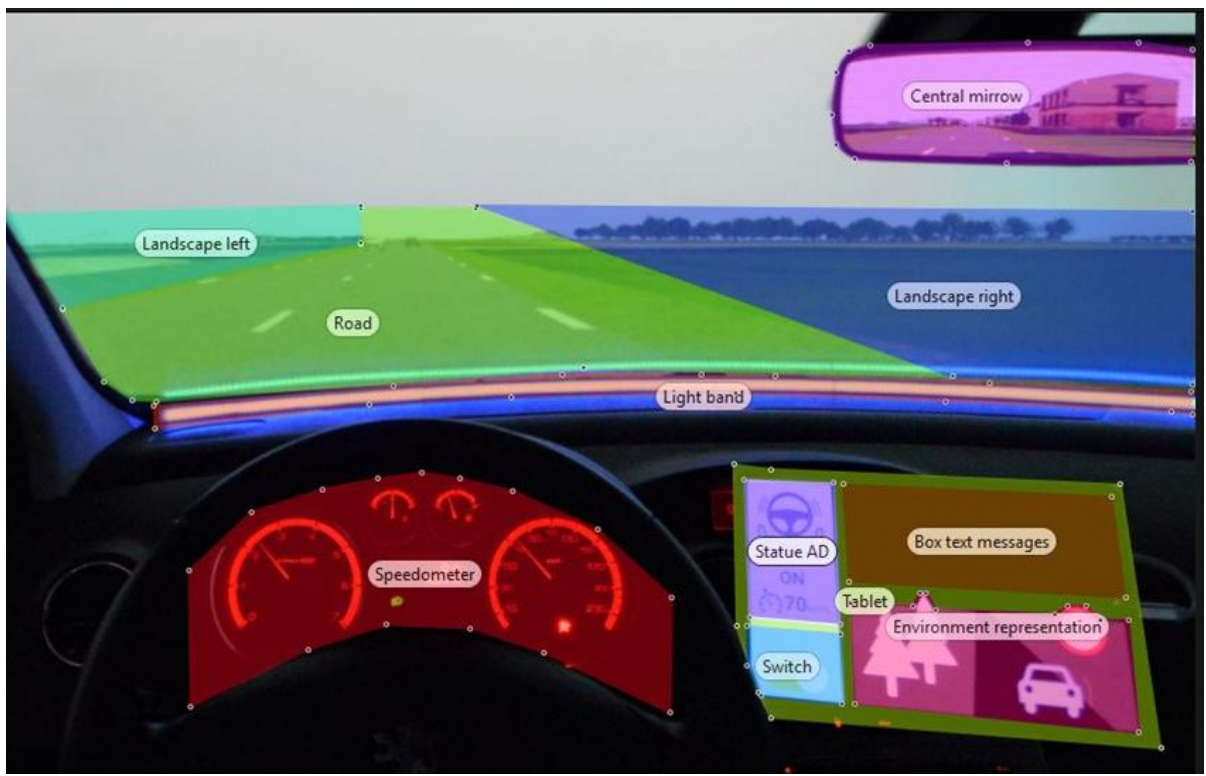

**Figure 2.** AOI used in this study for analyzing gaze behavior.

**Measures**

Overall, the study collected data from the following measurement devices, including eye-tracking, heart rate, electrodermal response, and wrist motility. However, our analysis focused primarily on three specific eye-tracking measures and self-reported trust, which are the local foci of the study's objectives.

*Eye tracking.* We considered three metrics for each participant: Total Fixation Duration, Fixation Count, and Time to First Fixation. A fixation is defined as a moment when the gaze remains relatively stable on a specific location, indicating focused visual attention (Martinez-Conde et al., 2004). While the measures for the speedometer and central mirror were directly obtained from the eye tracker software (Tobii), those for the environment and Tablet required additional data pre-processing to calculate, summarized in Table 1.

**Table 1.** Overview of AOI Measures and their calculations based on the zones as illustrated in Figure 4.

| Measure Definition | Data Pre-Processing | Study Interpretation |
|---|---|---|
| **Proportion of Fixation Duration:** The proportion of total time a 'participant's gaze remains fixed on a specific area within the visual environment. | ***Environment***: Calculated by summing the fixation times across the four designated environmental zones (the right and left landscape, road, and light band) and dividing the result by the Total Time of Interest Duration. ***Tablet***: Calculated by summing the fixation times across the four Tablet and green zones, then dividing the result by the Total Time of Interest Duration ***Speedometer:*** **Calculated** by dividing the fixation times of the speedometer into the Total Time of Interest Duration ***Central Mirror***: Calculated by dividing the fixation times of the central mirror into the Total Time of Interest Duration | **Allocation of visual attention:** These zones represent vital areas within the 'participant's field of view, and combining them allows us to assess the overall attention distribution across these critical regions of the visual environment. So, we can evaluate *how participants allocate their visual attention across the different parts of each factor: environment, Tablet, speedometer, and central mirror.* |
| **Proportion of Fixation Count:** The total number of individual fixations within a specific area of interest within the visual environment. | ***Environment***: Calculated the total fixation count by summing the fixations across four designated zones and dividing them into the Total Time of Interest Fixation Count. ***Tablet***: Calculated the total fixation count by summing the number of fixations across four Tablet and green zones, then dividing the result by the Total Time of Interest Fixation Count. ***Speedometer***: Calculated by dividing the fixation count of the speedometer into the Total Time of Interest Fixation Count ***Central Mirror***: Calculated by dividing the fixation count of the central mirror into the Total Time of Interest Fixation Count | **Distribution of visual attention:** By aggregating the fixation counts from these areas, we can assess how frequently 'participants' gaze shifts to these regions, providing insights into *how their visual attention is distributed across the four factors.* |
| **Time to First Fixation:** The time elapses from the beginning of a visual stimulus or scene until the participant's gaze fixates on a specific AOI. | ***Environment:*** Calculated the time to first fixation by identifying the shortest time it took participants to fixate on any of the four designated zones. ***Tablet***: Calculated the time to first fixation by identifying the shortest time participants had to fixate on any of the four Tablet and green zones. | **Priorities of visual attention:** By measuring the time participants fixate on each zone, we can evaluate *how quickly these areas attract attention, helping us understand the immediate visual priorities within the four factors.* |

*Self-reported trust in driving.* Participants took a survey after the driving task of each session, with seven Likert-scale, self-report measures of trust and expertise. Specifically for trust, the following question was asked at the end of the study: On a scale of 1 to 7, how would you rate your trust level in the system?

# DATA ANALYTICS AND RESULTS

## Generalized Linear Mixed Model (GLMM)

Prior to the analysis, the normality of the response variables was assessed using the Shapiro-Wilk test, which revealed significant deviations from normality for Fixation Duration ($FD$: $W = 0.814$, $p < .001$), Fixation Count ($FC$: $W = 0.869$, $p < .001$) and Time to First Fixation ($TFF$: $W = 0.562$, $p < .001$). Given the positively skewed nature of these continuous response variables and accounting for the nested structure of the experimental design, separate hierarchical Generalized Linear Mixed Models (GLMMs) were applied for FD, FC, and TFF. To address the skewed distribution, a Gamma distribution with a log-link function was used, as it is well-suited for positively skewed data and ensures that predicted values remain strictly positive (Nelder & Wedderburn, 1972).

These GLMMs were fitted using maximum likelihood estimation with Laplace Approximation (Ju et al., 2020). Firstly, to address potential issues arising from zero '0' values in the dataset, a small constant was added to all three response variables, creating shifted versions of $FD_{shift}$, $FC_{shift}$, and $TFF_{shift}$. These adjustments ensure compatibility with the Gamma distribution used in the GLMM, as the Gamma family requires strictly positive values for analysis. By applying this transformation, the integrity of the statistical assumptions underlying the model is maintained while preserving the meaningful variability in the original data. Each of the models was fitted using the `bobyqa` optimizer (Powell, 2009) with a maximum of $1 \, X \, 10^5$ iterations to ensure convergence. Analyses were conducted using the `glmer()` function from the `lme4` package (Bates et al., 2024) in RStudio.

### Fixation Duration (FD)

To predict $FD_{shift}$, the model included fixed effects as the nested structure that was explicitly modeled as Time (Linear) within modes ($m$: Automation, Handover, Manual, Takeover) and Modes nested within factors ($f$: Tablet, Environment, Speedometer, Central Mirror). At the start of the model-building steps, both linear and higher-order polynomial terms of time, i.e., quadratic and cubic terms, were initially considered; however, these higher-order terms were ultimately excluded from the model, because including them did not significantly improve the model's fit or provide additional explanation for variance in $FD_{shift}$ ($p > .05$). Additionally, random intercepts and slopes for Time (Linear) were included for each participant, accounting for individual variability in both baseline fixation duration and changes over time. So, the model is represented with fixed and random effects mathematically as:

$$\log\left(E[\text{FD}_{\text{Shift},ij} \mid X_{ij}]\right) = \beta_0 + \sum_f \beta_f f + \sum_{m(f)} \beta_{m(f)} m(f) + \sum_{t_1(m(f))} \beta_{t_1(m(f))} t_1(m(f)) + u_{0j} + u_{1j} t_1$$

(1)

In Formula 1, the term $E[FD_{Shift,ij} \mid X_{ij}]$ represents the expected value (mean) of $\text{FD}_{\text{Shift},ij}$ for observation $i$ of subject $j$, given the set of predictors ($X_{ij}$), which includes the fixed effects predictors for each factor ($\sum_f \beta_f f$), the fixed effects for mode ($m$) nested within $f$, ($\sum_{m(f)} \beta_{m(f)} m(f)$), and the effects of $t_1$(linear time nested within $m(f)$: $\sum_{t_1 m(f)} \beta_{t_1(m(f))} t_1(m(f))$). In addition, the model incorporates random effects component (($u_{0j} + u_{1j} t_1$)) to account for subject-specific deviations from the fixed effects, including a random intercept ($u_{0j}$), capturing variability in the baseline response for each subject ($j$), and a random slope ($u_{1j}$), allowing the effect of t₁ to vary across subjects. $u_{0j}$ and $u_{1j}$ are assumed to follow a multivariate normal distribution with a mean of zero and a variance-covariance structure ($N(0, \Sigma)$). Results were interpreted within this hierarchical nested framework rather than interaction to capture insights into how time effects vary across nested combinations of Factors and Modes (see Table 2).

Before the results summary of GLMM to predict $FD_{shift}$, an Intra-Class Correlation (ICC) was calculated to assess the proportion of variance attributable to the different levels in the hierarchical model, i.e., quantifying how much of the variability in $FD_{shift}$ is driven by individual differences (subject-level) versus differences due to the nested structure involving $f$, $m$, and $t_1$. The analysis revealed that the total ICC was 0.84, indicating that 83.8% of the total variance was explained by the combined effects of subject-level differences and the nested structure. The subject-level ICC was 0.018, suggesting that only 1.8% of the total variance was attributable to individual differences in subjects, as captured by the random intercept and slope for $t_1$. In contrast, the nested structure ICC was 0.82, indicating that 82% of the variance was explained by the fixed effects related to the hierarchical design. That is, the variability in $FD_{shift}$ is predominantly driven by the hierarchical structure rather than individual subject differences. Additionally, the model was not singular, confirming the appropriateness of the random effects specification and the partitioning of variance.

Random effects in GLMM were specified at the subject level with both an intercept and a slope for Time (Linear). The variance ($\sigma^2$) and the standard deviation (SD) of each of them is the random intercept: $\sigma^2 = 0.05$ ($SD = 0.22$), the random slope for Time (Linear): $\sigma^2 = 0.003$ ($SD = 0.05$) (the correlation between the intercept and slope was 0.36), and the residual: $\sigma_\epsilon^2 = 0.47$ ($SD = 0.69$). While scaled residuals ranged from -1.46 to 10.83, with a median of -0.13, indicating a slight positive skew but within acceptable limits for model diagnostics, model fit indices were as follows: AIC = 14,975.0, BIC = 15,183.4, Log-likelihood = -7451.5, and Deviance = 14,903.0. These results indicate that the nested structure of Time within Mode and Mode within Factor (Factor → Mode → Time) is appropriately modeled, with random variation accounted for at the subject level; the scaled residuals and fit indices support the adequacy of the model.

The results of fixed effects summarized in Table 2 indicate significant impacts within the nested levels of $FD_{shift}$, suggesting that $FD_{shift}$ was influenced by driving modes within each factor. The time-based effects within these nested levels were generally weaker. For instance, there were negligible changes in attention during automation mode on the Tablet ($\beta_{Std} = -0.003$, $p = .169$) and in the central mirror during takeover mode ($\beta_{Std} = -0.016$, $p = .271$). On the other hand, significant changes in attention were observed for the Tablet during manual mode ($p < .001$) and for the central mirror during Handover ($p = .002$).

As seen in Table 2, the significant findings of time-based effects (Factor ⟹ Mode ⟹ Time ($t_l$)) show that for the Tablet nested within Manual mode, with each one-unit increase in time, the proportion of fixation duration ($FD_{Shift}$) decreased by 0.016 standardized units ($\beta_{std} = -0.016$, $p < .001$). This decline suggests a gradual reduction in visual attention allocated to tablet-related areas over time (Figure 5). For the Speedometer nested within Manual mode, the standardized coefficient for the linear time component was similarly negative, with $FD_{Shift}$ decreasing by 0.006 standardized units for each one-unit increase in time ($\beta_{std}$ = -0.006, $p$ = .002). This decrease indicates that participants allocated less visual attention to the speedometer over time, possibly as participants grew more comfortable or familiar with manual mode and relied less on speed-related information.

**Table 2.** GLMM results of fixed effects for fixation duration regarding how fixation duration, a proxy for allocation of visual attention, are influenced by varying mode and factor combinations.

| | **Factor** | | | | | **Factor ⟹ Mode** | | | | **Factor ⟹ Mode ⟹ Time (t₁)** | | | |
|---|---|---|---|---|---|---|---|---|---|---|---|---|---|
| | *β (SE)* | *βStd* | *t-value* | *Pr(>|z|)* | | *β (SE)* | *βStd* | *t-value* | *Pr(>|z|)* | *β (SE)* | *βStd* | *t-value* | *Pr(>|z|)* |
| **Tablet** | -4.71 (0.09) | -0.076 | -53.26 | **< .001** | **Automation** | 0.68 (0.09) | 0.011 | 7.52 | **< .001** | -0.19 (0.14) | -0.003 | -1.38 | 0.169 |
| | | | | | **Handover** | 2.89 (0.09) | 0.046 | 31.47 | **< .001** | 0.10 (0.14) | 0.002 | 0.72 | 0.473 |
| | | | | | **Takeover** | 3.06 (0.09) | 0.049 | 32.54 | **< .001** | 0.07 (0.18) | 0.001 | 0.40 | 0.691 |
| | | | | | **Manual** | N/A | N/A | N/A | N/A | -0.98 (0.16) | -0.016 | -6.20 | **< .001** |
| **Speedometer** | -1.93 (0.07) | -0.031 | -28.76 | **< .001** | **Automation** | -1.24 (0.07) | -0.020 | -17.68 | **< .001** | 0.16 (0.14) | 0.003 | 1.15 | 0.250 |
| | | | | | **Handover** | -0.06 (0.09) | -0.001 | -0.68 | 0.497 | -0.13 (0.22) | -0.002 | -0.60 | 0.548 |
| | | | | | **Takeover** | 0.25 (0.08) | 0.004 | 3.19 | **0.001** | 0.13 (0.19) | 0.002 | 0.68 | 0.495 |
| | | | | | **Manual** | N/A | N/A | N/A | N/A | -0.37 (0.12) | -0.006 | -3.06 | **0.002** |
| **Central Mirror** | -4.91 (0.11) | -0.079 | -45.83 | **< .001** | **Automation** | 0.80 (0.11) | 0.013 | 6.98 | **< .001** | -0.31 (0.18) | -0.005 | -1.76 | 0.078 |
| | | | | | **Handover** | 1.83 (0.34) | 0.029 | 5.35 | **< .001** | -0.58 (0.66) | -0.009 | -0.88 | 0.379 |
| | | | | | **Takeover** | 1.38 (0.34) | 0.022 | 4.02 | **< .001** | -1.01 (0.92) | -0.016 | -1.10 | 0.271 |
| | | | | | **Manual** | N/A | N/A | N/A | N/A | -0.47 (0.26) | -0.008 | -1.80 | 0.071 |
| **Environment** | N/A | N/A | N/A | N/A | **Automation** | -0.05 (0.07) | -0.001 | -0.69 | 0.490 | -0.14 (0.12) | -0.002 | -1.14 | 0.253 |
| | | | | | **Handover** | -0.41 (0.07) | -0.007 | -6.24 | **< .001** | -0.11 (0.13) | -0.002 | -0.89 | 0.371 |
| | | | | | **Takeover** | -0.49 (0.07) | -0.008 | -6.92 | **< .001** | 0.01 (0.15) | 0.000 | 0.07 | 0.944 |
| | | | | | **Manual** | N/A | N/A | N/A | N/A | 0.07 (0.13) | 0.001 | 0.54 | 0.588 |

*Note. "β" and "β (SE)" refer to the unstandardized regression coefficient and its Standard Error, respectively, while "β" denotes the standardized regression coefficient. Time (Linear) within Mode and Mode within Factor specified in a symbolic frame of Factor ⟹ Mode ⟹ Time. "N/A" stands for Manual Mode, and "Environment" is used as a reference condition.*

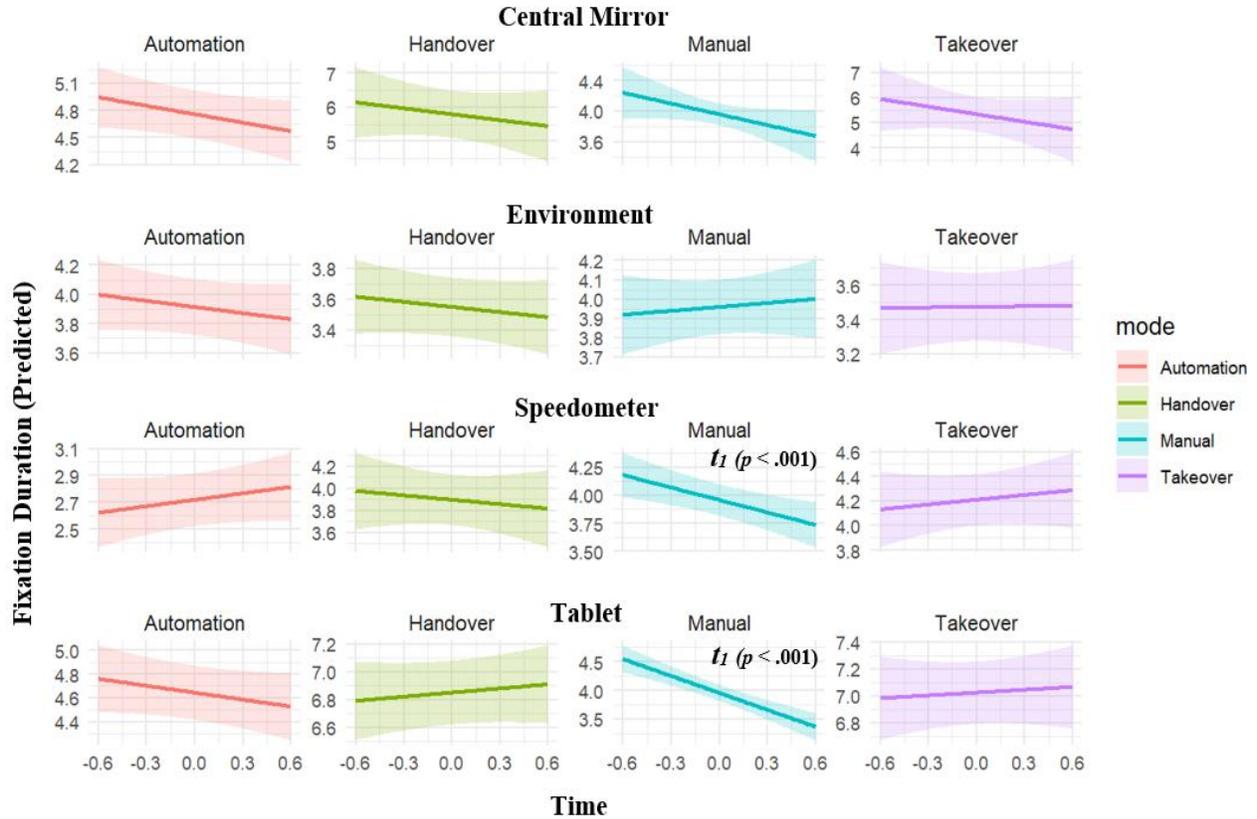

**Figure 3.** Predicted fixation duration over time (linear) across modes within factors. Time (Linear) is represented on the x-axis as standardized and centered units (i.e., deviations from the mean). Shaded areas indicate 95% Confidence Intervals (%95 CI), highlighting variations in attention across modes within factors.

Following the significant fixed effects of Mode within Factor (Factor → Mode), pairwise contrast analyses were conducted to compare the different modes within each factor, revealing how participants allocated their visual attention across the four factors during each mode, as summarized in Table 3. Here is the summary of significant mode effects within each factor. For the Environment factor, $FD_{Shift}$ was significantly higher in the Manual mode than in the Takeover ($p < .001$), while it did not differ between Manual and Automation ($p = .506$) or Handover ($p < .342$). Automation exhibited significantly higher $FD_{Shift}$ than both Handover and Takeover ($p < .001$), while no significant difference was observed between Handover and Takeover, $p = .353$. These results suggest that participants allocated more visual attention to the Environment during manual mode, reflecting the heightened visual demands in this phase.

For the Tablet factor, the Automation mode elicited significantly higher $FD_{Shift}$ than Manual and Handover ($p < .001$). Manual mode consistently and significantly had the lowest $FD_{Shift}$ compared to all other modes ($p < .001$), whereas Handover and Takeover were equal ($p = 1.000$). For the Speedometer, the Automation mode had a significantly lower $FD_{Shift}$ than Manual, Handover, and Takeover ($p < .001$), whereas all three were equal $FD_{Shift}$. These results suggest that participants allocated significantly more attention to the Speedometer during Automation mode, while the other modes showed relatively similar levels of attention. For the Central Mirror factor, $FD_{Shift}$ was significantly lower in Manual mode compared to Automation, Handover, and Takeover ($p < .001$). Additionally, Automation had a significantly lower $FD_{Shift}$ compared to Handover ($p = .001$) and marginally lower than Takeover ($p = .081$), with no difference observed between Handover and Takeover ($p = 1.000$). These findings suggest that participants focused more on the Central Mirror during the control-oriented transition modes (i.e., handover and takeover), reflecting the importance of situational awareness in active control.

Overall, mode-based effects demonstrate that transitional modes, e.g., Takeover mode, drew heightened visual attention to the Central Mirror, while Automation elicited increased attention toward the technological aspects of driving (e.g., Tablet or Speedometer). In contrast, the Manual mode directed participants' attention more prominently toward the Environment, reflecting the heightened visual demands of active control.

**Table 3**. Pairwise contrast analysis of $FD_{shift}$ across modes nested within factors in the hierarchical structure.

| Factor | Contrast *(Numerator / Denominator)* | $\theta$ | $SE\ \theta$ | $\beta_{Std}$ | *z-test* | *p-value* |
|---|---|---|---|---|---|---|
| Environment | Manual / Automation | 0.88 | 0.04 | 2.90 | -2.71 | 0.506 |
| | Manual / Handover | 1.14 | 0.05 | 3.76 | 2.93 | 0.342 |
| | Manual / Takeover | 1.31 | 0.06 | 4.32 | 5.69 | 0.000 |
| | Automation / Handover | 1.29 | 0.06 | 4.26 | 5.65 | 0.000 |
| | Automation / Takeover | 1.49 | 0.07 | 4.89 | 8.28 | 0.000 |
| | Handover / Takeover | 1.15 | 0.06 | 3.78 | 2.91 | 0.353 |
| Tablet | Manual / Automation | 0.62 | 0.04 | 2.04 | -7.71 | 0.000 |
| | Manual / Handover | 0.12 | 0.01 | 0.39 | -34.07 | 0.000 |
| | Manual / Takeover | 0.11 | 0.01 | 0.38 | -33.98 | 0.000 |
| | Automation / Handover | 0.19 | 0.01 | 0.63 | -32.10 | 0.000 |
| | Automation / Takeover | 0.18 | 0.01 | 0.61 | -31.92 | 0.000 |
| | Handover / Takeover | 0.97 | 0.05 | 3.18 | -0.60 | 1.000 |
| Speedometer | Manual / Automation | 3.02 | 0.15 | 9.94 | 22.91 | 0.000 |
| | Manual / Handover | 0.96 | 0.06 | 3.17 | -0.59 | 1.000 |
| | Manual / Takeover | 0.87 | 0.05 | 2.88 | -2.51 | 0.671 |

| | | θ | SE θ | Mean Ratio | β | p |
|---|---|---|---|---|---|---|
| | Automation / Handover | 0.32 | 0.02 | 1.05 | -17.90 | 0.000 |
| | Automation / Takeover | 0.29 | 0.02 | 0.95 | -22.49 | 0.000 |
| | Handover / Takeover | 0.91 | 0.06 | 2.98 | -1.43 | 0.999 |
| Central Mirror | Manual / Automation | 0.63 | 0.05 | 2.08 | -5.79 | 0.000 |
| | Manual / Handover | 0.22 | 0.05 | 0.72 | -6.53 | 0.000 |
| | Manual / Takeover | 0.28 | 0.07 | 0.94 | -5.37 | 0.000 |
| | Automation / Handover | 0.35 | 0.08 | 1.14 | -4.67 | 0.001 |
| | Automation / Takeover | 0.45 | 0.10 | 1.48 | -3.49 | 0.081 |
| | Handover / Takeover | 1.30 | 0.41 | 4.28 | 0.83 | 1.000 |

*Note.* "$\theta$" and "SE $\theta$" refer to the Ratio of two group means, "Mean Ratio" and its Standard Error, respectively, while "$\beta$" denotes the standardized contrast coefficient. If $\theta > 1$, the numerator has a higher $FD_{Shift}$ than the denominator.

### Fixation Count (FC)

To predict $FC_{Shift}$ while accounting for the nested structure of the experimental design, we applied a second hierarchical GLMM. During the model-building process, we initially considered both linear and higher-order polynomial terms of time, specifically quadratic and cubic terms. We included linear ($t_1$) and quadratic ($t_2$) terms as fixed effects nested within the mode, with the mode further nested within the factors. However, we did not include the cubic term, as its inclusion did not significantly improve the model's fit or provide additional explanation for the variance in $FC_{shift}$, $p > .05$. Additionally, due to a singularity issue in the initial model, the random-effects structure was simplified by removing the random slopes for Time as $t_1$ and $t_2$, retaining only the random intercepts for participants in the analysis. The model uses a Gamma distribution for the response variable with a log-link function, which ensures predicted values remain strictly positive.

Prior to predicting $FC_{shift}$, an ICC was calculated to assess the proportion of variance attributable to the different levels in the hierarchical model, i.e., quantifying how much of the variability in $FC_{shift}$ is driven by individual differences (subject-level) versus differences due to the nested structure involving $f$, $m$, and $t_1$, and $t_2$. The analysis revealed that the total ICC was 0.847, indicating that 84.7% of the total variance was explained by the combined effects of subject-level differences and the nested structure. The subject-level ICC was 0.0033, suggesting that only 0.33% of the total variance was attributable to individual differences in subjects, as captured by the random intercept. In contrast, the nested structure ICC was 0.844, indicating that 84.4% of the variance was explained by the fixed effects related to the hierarchical design. The variability in $FC_{shift}$ is pretty much driven by the hierarchical structure rather than individual subject differences. Additionally, the model was not singular, confirming the appropriateness of the random effects specification and the partitioning of variance. Accordingly, the model is represented mathematically in Formula 3:

$$\log\left(E[\text{FC}_{\text{Shift},ij} \mid X_{ij}]\right) = \beta_0 + \sum_f \beta_f f + \sum_{m(f)} \beta_{m(f)} m(f) + \sum_{t_1(m(f))} \beta_{t_1(m(f))} t_1(m(f)) + \sum_{t_2(m(f))} \beta_{t_2(m(f))} t_2(m(f)) + u_{0j}$$

(3)

Where $Y_{ij}$ represents $FC_{ij}$ for observation $i$ of participant $j$. Unlike in Formula 1, $\beta_{t_2(m(f))} t_2\big(m(f)\big)$ represents the effect of quadratic time (i.e., nonlinear trend) nested within mode and factor; that is, $\beta_{t_2(m(f))}$ quantifies how curvilinear changes in time (quadratic) influence the expected $FC_{shift}$ within each nested mode within each factor. The random intercept $u_{0j}$ follows a normal distribution with a mean of zero '0' and a variance ($\sigma^2$) to account for between-subject differences.

Random effects ($u_{0j} = \sim N(0,\ \sigma^2)$) in the GLMM were specified at the subject level with a random intercept. The variance ($\sigma^2$) and standard deviation ($SD$) for the random intercept were $\sigma^2 = 0.01$ ($SD = 0.07$). The residual variance was $\sigma_\in^2 = 0.23$ ($SD = 0.48$). Scaled residuals ranged from -2.05 to 7.49, with a median of -0.07, indicating a slight positive skew but remaining within acceptable limits for model diagnostics. Model fit indices were as follows: AIC = -4464.1, BIC = -4174.7, Log-likelihood = 2282.0, and Deviance = 4564.1. Overall, the results of the random effects indicate that the random variation at the subject level is appropriately accounted for, and the scaled residuals and fit indices support the adequacy of the model.

$FC_{shift}$ results in Table 4 show significant impacts within the nested levels of modes within the factors, suggesting that $FC_{shift}$ was influenced by various driving modes within each factor. The notable time-based effects were observed in the tablet during the handover and manual modes. Accordingly, for time (quadratic) within the manual mode nested within the Tablet (Tablet → Manual → Time ($t_2$); see Table 4) relative to the reference context (manual mode within the environment) was statistically significant, $p <$ .001.

The significant linear ($t_1$) time effect for the Manual mode nested within Tablet indicates that the proportion of $FD_{Shift}$ decreased by 0.39 standardized units with each one-unit increase in time ($\beta_{Std}$ = -0.391, $p$ = .012), reflecting declining visual priority for tablet as the task progressed. For the quadratic time component ($t_2$), the significant positive effect ($\beta_{Std}$ = 0.79, $p <$ .001) suggests a non-linear trend; that is, initially, visual attention decreased (quicker shifts away from the tablet), but later increased (slower attention shifts toward the tablet), highlighting dynamic adjustments in attention prioritization over time in manual mode (See Figure 6). For the Handover mode within Tablet, a significant positive quadratic time effect ($\beta_{Std}$ = 0.270, $p$ = .028) indicates a non-linear pattern of visual attention shifts. Initially, participants' fixation count increased toward tablet-related areas (slower attention shifts), but later decreased (quicker shifts away), reflecting a dynamic adjustment in visual priorities during the handover process.

The final GLMM model for predicting $FC_{Shift}$ presents the estimated fixed effect coefficients (standardized) for the factors, modes, and time, as shown in Table 4. Additionally, the results of the simple slope analysis are included. The random effects are represented as participant-specific deviations from the fixed effects, as illustrated in the following formula:

$$\log(E[\text{FC}_{\text{Shift},ij} \mid X_{ij}]) = -0.437 + (-3.981)f_{\text{Tablet}} + (-1.614)f_{\text{Speedometer}} + (-4.156)f_{\text{Central Mirror}} + (0.148)m_{\text{Automation}}(f_{\text{Environment}})$$
$$+ (-0.391)t_1(m_{\text{Manual}}(f_{\text{Tablet}})) + (0.785)t_2(m_{\text{Manual}}(f_{\text{Tablet}})) + (0.270)t_2(m_{\text{Handover}}(f_{\text{Tablet}})) + u_{0j}.$$

(4)

**Table 4**. GLMM results of fixed effects for fixation count illustrate how visual attention is distributed across the four factors during varying modes over time.

*Note. "β" and "β (SE)" refer to the unstandardized regression coefficient and its Standard Error, respectively, while "β" denotes the standardized regression*

| Factor | β (SE) | βStd | t-value | Pr(>\|z\|) | Mode | β (SE) | βStd | t-value | Pr(>\|z\|) | β (SE) | βStd | t-value | Pr(>\|z\|) | β (SE) | βStd | t-value | Pr(>\|z\|) |
|---|---|---|---|---|---|---|---|---|---|---|---|---|---|---|---|---|---|
| | | Factor | | | | Factor ⟹ Mode | | | | Factor ⟹ Mode ⟹ Time (Linear) | | | | Factor ⟹ Mode ⟹ Time (Quadratic) | | | |
| **Tablet** | -3.28 (0.06) | -3.981 | -53.77 | **< .001** | Automation | 0.56 (0.06) | 0.679 | 8.93 | **< .001** | -0.10 (0.09) | -0.122 | -1.08 | 0.278 | -0.14 (0.09) | -0.175 | -1.56 | 0.120 |
| | | | | | Handover | 2.22 (0.06) | 2.688 | 35.02 | **< .001** | 0.06 (0.10) | 0.076 | 0.65 | 0.515 | 0.22 (0.10) | 0.270 | 2.20 | **0.028** |
| | | | | | Takeover | 2.22 (0.07) | 2.687 | 32.59 | **< .001** | -0.02 (0.14) | -0.028 | -0.16 | 0.871 | -0.26 (0.14) | -0.310 | -1.82 | 0.069 |
| | | | | | Manual | N/A | N/A | N/A | N/A | -0.32 (0.13) | -0.391 | -2.54 | **0.012** | 0.65 (0.13) | 0.785 | 5.07 | **< .001** |
| **Speedometer** | -1.33 (0.05) | -1.614 | -29.09 | **< .001** | Automation | -1.10 (0.05) | -1.334 | -23.00 | **< .001** | 0.02 (0.09) | 0.030 | 0.27 | 0.789 | -0.10 (0.09) | -0.124 | -1.14 | 0.255 |
| | | | | | Handover | 0.04 (0.06) | 0.051 | 0.68 | 0.498 | -0.07 (0.14) | -0.080 | -0.46 | 0.648 | 0.10 (0.14) | 0.117 | 0.67 | 0.504 |
| | | | | | Takeover | 0.10 (0.06) | 0.120 | 1.70 | 0.089 | -0.05 (0.16) | -0.061 | -0.32 | 0.751 | -0.30 (0.16) | -0.362 | -1.86 | 0.063 |
| | | | | | Manual | N/A | N/A | N/A | N/A | -0.17 (0.08) | -0.201 | -1.96 | 0.051 | 0.02 (0.08) | 0.019 | 0.19 | 0.851 |
| **Central Mirror** | -3.43 (0.07) | -4.156 | -47.06 | **< .001** | Automation | 0.47 (0.08) | 0.565 | 5.97 | **< .001** | -0.17 (0.11) | -0.206 | -1.54 | 0.124 | -0.16 (0.11) | -0.190 | -1.41 | 0.157 |
| | | | | | Handover | 1.63 (0.28) | 1.973 | 5.82 | **< .001** | 0.10 (0.48) | 0.127 | 0.22 | 0.827 | -0.48 (0.61) | -0.582 | -0.79 | 0.427 |
| | | | | | Takeover | 1.25 (0.26) | 1.517 | 4.89 | **< .001** | -0.50 (0.80) | -0.603 | -0.62 | 0.536 | 0.14 (0.76) | 0.173 | 0.19 | 0.852 |
| | | | | | Manual | N/A | N/A | N/A | N/A | -0.20 (0.17) | -0.248 | -1.20 | 0.232 | -0.03 (0.18) | -0.037 | -0.17 | 0.864 |
| **Environment** | N/A | N/A | N/A | N/A | Automation | 0.12 (0.05) | 0.148 | 2.69 | **0.007** | -0.02 (0.08) | -0.022 | -0.22 | 0.828 | 0.00 (0.08) | -0.002 | -0.02 | 0.982 |
| | | | | | Handover | -0.14 (0.05) | -0.164 | -2.99 | **0.003** | 0.02 (0.08) | 0.022 | 0.21 | 0.833 | -0.04 (0.08) | -0.047 | -0.47 | 0.637 |
| | | | | | Takeover | -0.25 (0.05) | -0.301 | -4.84 | **< .001** | 0.13 (0.13) | 0.155 | 1.00 | 0.318 | 0.19 (0.13) | 0.232 | 1.53 | 0.127 |
| | | | | | Manual | N/A | N/A | N/A | N/A | 0.06 (0.08) | 0.072 | 0.71 | 0.480 | -0.01 (0.08) | -0.007 | -0.07 | 0.947 |

*coefficient—Time within Mode and Mode within Factor specified in a symbolic frame of Factor ⟹ Mode ⟹ Time. "N/A" stands for Manual Mode, and "Environment" is used as reference conditions.*

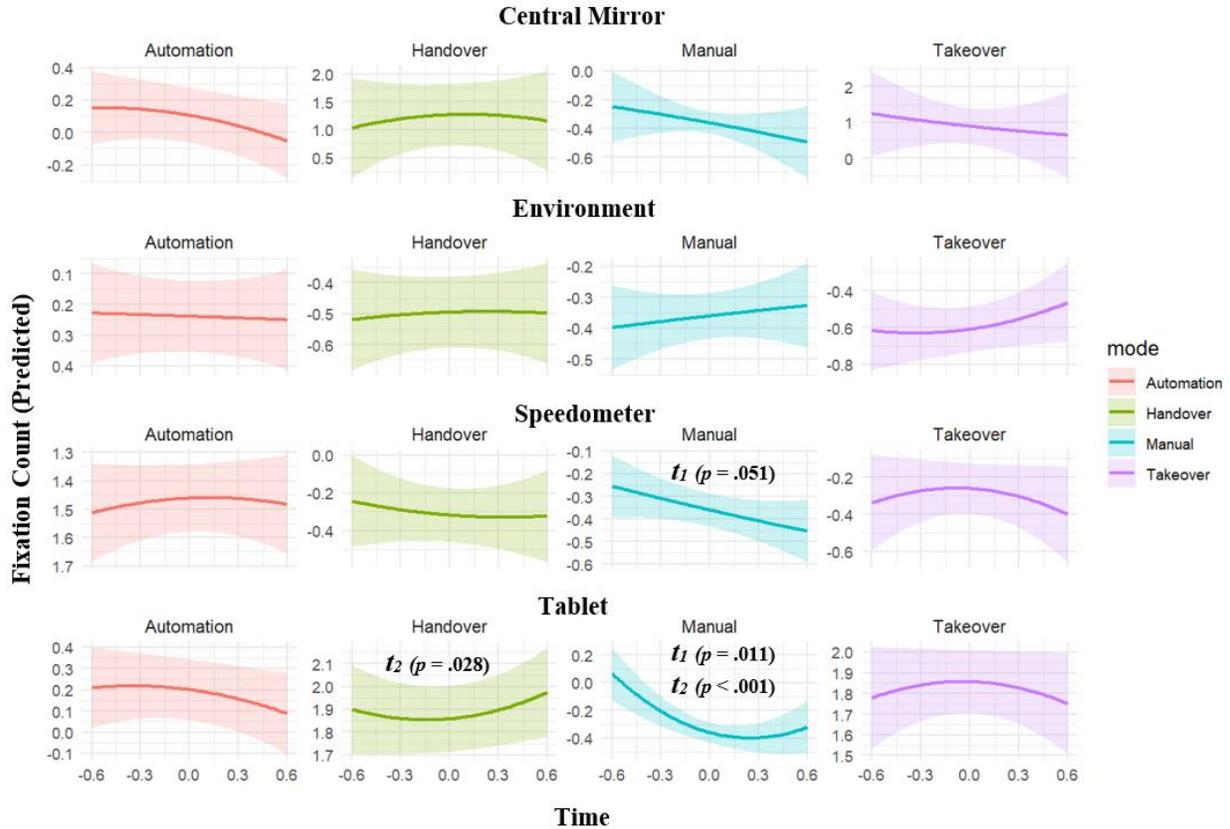

**Figure 4.** Predicted $FC_{Shifted}$ over time across modes within factors. Time (Quadratic) is represented on the x-axis as standardized and centered units (i.e., deviations from the mean). Shaded areas indicate %95 CI, highlighting variations in attention across modes within factors.

Following the significant fixed effects of mode within the Factor (Factor → Mode), pairwise contrast analyses were conducted to compare the different modes within each factor, revealing how participants shifted their gaze between these modes and highlighting variations in visual attention patterns, as summarized in Table 5. For the Environment factor, $FC_{Shift}$ was significantly lower in Takeover compared to Manual ($p < .001$) but did not differ from Automation ($p = .488$) or Handover ($p = .332$). Additionally, $FC_{Shift}$ was significantly higher in Handover and Takeover compared to Automation ($p < .001$), with no difference between Handover and Takeover ($p = .743$). These results indicate that gaze shifts in the Environment factor occur more frequently during Handover and Takeover modes, reflecting heightened visual attention in transitional phases. For the Tablet factor, $FC_{Shift}$ was significantly lower in Manual mode compared to the other three modes ($p < .001$), while Handover and Takeover exhibited higher $FC_{Shift}$ than Automation ($p < .001$) but were equal to each other ($p = 1.000$). The distribution of visual attention to the Tablet factor, based on fixation counts, shows that the Manual mode had significantly fewer fixations compared to the other modes, while Handover and Takeover involved more frequent fixations than Automation, reflecting heightened monitoring demands during transitional phases.

For the Speedometer factor, Automation had significantly lower $FC_{Shift}$ compared to Manual, Handover, and Takeover ($p < .001$), while the latter three modes exhibited similar $FC_{Shift}$ ($p > .05$). The distribution of visual attention for the Speedometer factor suggests that Automation mode required less frequent monitoring of speed, while Manual, Handover, and Takeover demanded consistent attention to speed-related information. For the Central Mirror factor, $FC_{Shift}$ was significantly lower in Manual mode compared to Automation, Handover, and Takeover ($p < .001$). Automation mode also had significantly lower $FC_{Shift}$ than Handover ($p = .007$), while no significant differences were observed between Automation and Takeover ($p = .139$) or between Handover and Takeover ($p = 1.000$). These results suggest that the distribution of visual attention shows minimal gaze shifts to the Central Mirror during Manual mode, with increased shifts during transitional modes, highlighting their monitoring demands.

The overall mode-based effects reveal that visual attention allocation varied significantly across modes and factors, with transitional modes (Handover and Takeover) consistently involving more frequent gaze shifts, particularly to the Environment and Tablet, reflecting heightened monitoring demands. Manual mode showed reduced gaze shifts overall, particularly to the Tablet and Central Mirror, indicating a focus on immediate task demands. In contrast, the Automation mode exhibited fewer gaze shifts to the Speedometer and Central Mirror, highlighting reduced reliance on these areas during automated control.

**Table 5**. Pairwise contrast analysis of $FC_{shift}$ across modes nested within factors in the hierarchical structure.

| Factor | Contrast (Numerator / Denominator) | $\theta$ | SE $\theta$ | $\beta_{Std}$ | z-test | p-value |
|---|---|---|---|---|---|---|
| **Environment** | Manual / Automation | 0.88 | 0.04 | 2.90 | -2.73 | 0.488 |
| | Manual/ Handover | 1.14 | 0.05 | 3.76 | 2.94 | 0.332 |
| | Manual/ Takeover | 1.29 | 0.06 | 4.24 | 5.08 | **< .001** |
| | Automation / Handover | 1.29 | 0.06 | 4.25 | 5.69 | **< .001** |

| | | θ | SE θ | | β_Std | p |
|---|---|---|---|---|---|---|
| | Automation / Takeover | 1.46 | 0.07 | 4.79 | 7.59 | **< .001** |
| | Handover / Takeover | 1.13 | 0.06 | 3.71 | 2.41 | 0.743 |
| **Tablet** | Manual / Automation | 0.57 | 0.04 | 1.87 | -9.00 | **< .001** |
| | Manual/ Handover | 0.11 | 0.01 | 0.36 | -35.02 | **< .001** |
| | Manual/ Takeover | 0.11 | 0.01 | 0.36 | -33.45 | **< .001** |
| | Automation / Handover | 0.19 | 0.01 | 0.63 | -32.13 | **< .001** |
| | Automation / Takeover | 0.19 | 0.01 | 0.63 | -30.04 | **< .001** |
| | Handover / Takeover | 0.99 | 0.06 | 3.26 | -0.13 | 1.000 |
| **Speedometer** | Manual / Automation | 3.02 | 0.14 | 9.93 | 23.07 | **< .001** |
| | Manual/ Handover | 0.96 | 0.06 | 3.16 | -0.62 | 1.000 |
| | Manual/ Takeover | 0.91 | 0.05 | 2.98 | -1.77 | 0.987 |
| | Automation / Handover | 0.32 | 0.02 | 1.05 | -18.04 | **< .001** |
| | Automation / Takeover | 0.30 | 0.02 | 0.99 | -20.95 | **< .001** |
| | Handover / Takeover | 0.94 | 0.07 | 3.10 | -0.87 | 1.000 |
| **Central Mirror** | Manual / Automation | 0.63 | 0.05 | 2.06 | -5.93 | **< .001** |
| | Manual / Handover | 0.20 | 0.06 | 0.65 | -5.76 | **< .001** |
| | Manual / Takeover | 0.28 | 0.07 | 0.93 | -5.14 | **< .001** |
| | Automation / Handover | 0.31 | 0.09 | 1.03 | -4.17 | **0.007** |
| | Automation / Takeover | 0.45 | 0.11 | 1.49 | -3.30 | 0.139 |
| | Handover / Takeover | 1.44 | 0.52 | 4.74 | 1.01 | 1.000 |

*Note.* *"θ" and "SE θ" refer to the Ratio of two group means, "Mean Ratio" and its Standard Error, respectively, while "$\beta_{Std}$" denotes the standardized contrast coefficient. If $\theta > 1$, the numerator has a higher $FC_{Shift}$ than the denominator.*

### Time to First Fixation (TFF)

The final GLMM for predicting $TFF_{Shift}$ follows the same structure as depicted in Formula 3. This model includes linear ($t_1$) and quadratic ($t_2$) time components nested within the mode, which is further nested within factors. The model retains only random intercepts for participants, capturing individual differences in baseline fixation times. Consistent with the previous specification, a Gamma distribution and a log-link function are used for the response variable, ensuring that predicted values remain strictly positive.

First, ICCs were calculated to assess the proportion of variance in time to first fixation attributable to the nested structure and subject-level differences. The total ICC for the combined effects of subjects and the nested structure was ICC = 0.5496, indicating that 54.96% of the variance in time to first fixation is attributable to differences across subjects and the nested factor structure. Further decomposition revealed that the subject-level ICC (random intercept) was ICC = 0.053, suggesting that 5.31% of the total variance is attributed to differences between subjects. The ICC for the nested structure time (linear and quadratic) within mode and mode within factor was ICC = 0.497, signifying that the unique contextual differences in the hierarchical structure of factors, modes, and time components explain approximately 50% of the variance in $TFF_{Shift}$.

Random effects ($u_{0j} = \sim N(0, \sigma^2)$) in the GLMM were specified at the subject level with a random intercept with the $\sigma^2 = 0.28$ ($SD = 0.53$) and the residual variance, $\sigma_\epsilon^2 = 2.38$ ($SD = 1.54$). Scaled residuals ranged from -0.65 to 11.65, with a median of -0.26, indicating a slight positive skew, but remaining within acceptable limits for model diagnostics. Model fit indices were as follows: AIC = 9404.0, BIC = 9693.5, Log-likelihood = -4652.0, and Deviance = 9304.0. Overall, the results of the random effects indicate that the random variation at the subject level is appropriately accounted for, and the scaled residuals and fit indices support the adequacy of the model.

With that in mind, Formula 5 builds on the hierarchical structure introduced in Formula 3, incorporating the same components to account for the nested design of the study. It aligns with the nested design of the study and highlights how multiple levels of predictors and random subject-level variation influence visual attention priorities (as measured by $TFF_{Shift}$):

$$\log(E[\text{TFF}_{\text{Shift},ij} \mid X_{ij}]) = \beta_0 + \sum_f \beta_f f + \sum_{m(f)} \beta_{m(f)} m(f) + \sum_{t_1(m(f))} \beta_{t_1(m(f))} t_1(m(f)) + \sum_{t_2(m(f))} \beta_{t_2(m(f))} t_2(m(f)) + u_{0j}$$
(5)

Table 6 shows fixed effects results of GLMM predicting $TFF_{Shift}$, where smaller mean values indicate quicker shifts in attention. In general, the nested levels of modes within the factors suggesting modes significantly influence attention patterns, especially during the Handover and Takeover phases, where attention tends to decline. As seen in Table 6 and Figure 6, the time-based trends (both $t_1$ and $t_2$) vary across modes within these factors, with notable effects observed specifically for Manual and Automation Mode in the factors of Tablet, Speedometer, and Environment. Accordingly, smaller $TFF_{Shift}$, values indicate quicker attention and higher visual priority, while larger values indicate slower attention and lower priority. The results of time effects are interpreted as nested effects rather than interactions to reflect the hierarchical structure of the data.

The significant findings show that for the Tablet factor nested within Manual mode, the linear time component indicates that with each one-unit increase in time, the TFF increased by 0.028 standardized units ($\beta_{Std}$ = .028, $p$ < .001). This suggests a gradual decrease in visual priority for Tablet-related areas over time. The significant quadratic component showed a negative trend, reflecting a non-linear shift where TFF initially decreased (quicker attention) and later increased (slower attention) ($\beta_{Std}$ = -.029, $p$ < .001). This pattern highlights how the prioritization of Tablet-related stimuli changes dynamically within Manual mode.

For the Environment factor, nested within Automation mode, the standardized coefficient for the linear time component was positive ($\beta_{Std}$ = 0.014 units per one-unit increase in time), indicating a steady decline in visual priority for Environment-related areas ($p$ = .003). Conversely, within Manual mode, a negative linear effect showed that $TFF_{Shif}$ decreased by 0.015 units with each one-unit increase in time, suggesting an increasing visual priority for Environment-related areas as time progressed ($p$ = .001). Additionally, the positive quadratic time component for the Environment within Manual mode revealed a non-linear trend, with $TFF_{Shif}$ initially increasing (slower attention) before decreasing (quicker attention) ($\beta_{Std}$ = 0.009, $p$ = .043). This underscores the dynamic adjustments in visual attention priorities over time. Consequently, using the standardized coefficients and focusing on the significant effects from Table 6, the formula for the GLMM results of $TFF_{Shift}$ is updated based on the results as follows:

$$\log(E[\text{TFF}_{\text{Shift},ij} \mid X_{ij}]) = 0.002 + 0.063 f_{\text{Tablet}} + 0.021 f_{\text{Speedometer}} + 0.07 f_{\text{Central Mirror}}$$

$$-0.013 m_{\text{Automation}}(f_{\text{Environment}}) + 0.042 m_{\text{Automation}}(f_{\text{Speedometer}}) - 0.016 m_{\text{Handover}}(f_{\text{Environment}})$$

$$-0.049 m_{\text{Handover}}(f_{\text{Tablet}}) - 0.014 m_{\text{Handover}}(f_{\text{Speedometer}}) - 0.048 m_{\text{Handover}}(f_{\text{Central Mirror}})$$

$$-0.015 m_{\text{Takeover}}(f_{\text{Environment}}) - 0.052 m_{\text{Takeover}}(f_{\text{Tablet}}) + 0.007 m_{\text{Takeover}}(f_{\text{Speedometer}})$$

$$-0.06 m_{\text{Takeover}}(f_{\text{Central Mirror}}) - 0.015 t_1(f_{\text{Environment}} : m_{\text{Manual}}) + 0.028 t_1(f_{\text{Tablet}} : m_{\text{Manual}})$$

$$+0.014 t_1(f_{\text{Speedometer}} : m_{\text{Manual}}) + 0.014 t_1(f_{\text{Environment}} : m_{\text{Automation}}) + 0.009 t_2(f_{\text{Environment}} : m_{\text{Manual}})$$

$$-0.029 t_2(f_{\text{Tablet}} : m_{\text{Manual}}) + u_{0j}$$

$$(6)$$

**Table 6**. GLMM results of fixed effects for $TFF_{Shift}$ about how visual attention priorities are distributed across the four factors during varying modes over time.

*Note. "β" and "β (SE)" refer to the unstandardized regression coefficient and its Standard Error, respectively, while "β" denotes the standardized regression*

| Factor | β (SE) | βStd | t-value | Pr(>\|z\|) | Mode | β (SE) | βStd | t-value | Pr(>\|z\|) | β (SE) | βStd | t-value | Pr(>\|z\|) | β (SE) | βStd | t-value | Pr(>\|z\|) |
|---|---|---|---|---|---|---|---|---|---|---|---|---|---|---|---|---|---|
| | | | | | | **Factor ⟹ Mode** | | | | **Factor ⟹ Mode ⟹ Time (Linear)** | | | | **Factor ⟹ Mode ⟹ Time (Quadratic)** | | | |
| **Tablet** | 3.29 (0.17) | 0.063 | 19.12 | **<.001** | Automation | 0.05 (0.18) | 0.001 | 0.28 | 0.778 | -0.06 (0.26) | -0.001 | -0.25 | 0.805 | 0.27 (0.25) | 0.005 | 1.11 | 0.268 |
| | | | | | Handover | -2.59 (0.18) | -0.049 | -14.71 | **<.001** | -0.04 (0.27) | -0.001 | -0.15 | 0.885 | -0.18 (0.27) | -0.003 | -0.66 | 0.508 |
| | | | | | Takeover | -2.73 (0.19) | -0.052 | -14.31 | **<.001** | 0.12 (0.41) | 0.002 | 0.30 | 0.766 | 0.32 (0.39) | 0.006 | 0.81 | 0.419 |
| | | | | | Manual | N/A | N/A | N/A | N/A | 1.46 (0.34) | 0.028 | 4.23 | **<.001** | -1.55 (0.37) | -0.029 | -4.22 | **<.001** |
| **Speedometer** | 1.08 (0.13) | 0.021 | 8.14 | **<.001** | Automation | 2.20 (0.14) | 0.042 | 16.13 | **<.001** | 0.06 (0.26) | 0.001 | 0.23 | 0.818 | 0.09 (0.24) | 0.002 | 0.37 | 0.715 |
| | | | | | Handover | -0.73 (0.18) | -0.014 | -4.09 | **<.001** | -0.04 (0.27) | -0.001 | -0.15 | 0.885 | 0.21 (0.39) | 0.004 | 0.54 | 0.589 |
| | | | | | Takeover | 0.37 (0.16) | 0.007 | 2.27 | **0.023** | -0.04 (0.43) | -0.001 | -0.09 | 0.928 | 0.26 (0.42) | 0.005 | 0.62 | 0.535 |
| | | | | | Manual | N/A | N/A | N/A | N/A | 0.71 (0.25) | 0.014 | 2.86 | **0.004** | 0.47 (0.25) | 0.009 | 1.85 | 0.065 |
| **Central Mirror** | 3.65 (0.20) | 0.07 | 17.93 | **<.001** | Automation | -0.14 (0.22) | -0.003 | -0.64 | 0.523 | 0.09 (0.31) | 0.002 | 0.32 | 0.751 | 0.20 (0.31) | 0.004 | 0.64 | 0.522 |
| | | | | | Handover | -2.50 (0.78) | -0.048 | -3.20 | **.001** | -1.47 (1.30) | -0.028 | -1.13 | 0.259 | -1.62 (1.68) | -0.031 | -0.96 | 0.335 |
| | | | | | Takeover | -3.14 (0.68) | 0.060 | -4.59 | **<.001** | -2.19 (1.89) | -0.042 | -1.16 | 0.248 | -1.81 (1.99) | -0.034 | -0.91 | 0.365 |
| | | | | | Manual | N/A | N/A | N/A | N/A | 0.65 (0.48) | 0.012 | 1.35 | 0.176 | -0.13 (0.51) | -0.003 | -0.26 | 0.795 |
| **Environment** | N/A | N/A | N/A | N/A | Automation | -0.70 (0.13) | -0.013 | -5.54 | **<.001** | 0.72 (0.24) | 0.014 | 2.99 | **0.003** | 0.25 (0.24) | 0.005 | 1.06 | 0.292 |
| | | | | | Handover | -0.85 (0.13) | -0.016 | -6.59 | **<.001** | 0.04 (0.24) | 0.001 | 0.14 | 0.886 | 0.24 (0.23) | 0.005 | 1.02 | 0.306 |
| | | | | | Takeover | -0.77 (0.15) | -0.015 | -5.14 | **<.001** | -0.48 (0.41) | -0.009 | -1.16 | 0.248 | -0.56 (0.38) | -0.011 | -1.50 | 0.135 |
| | | | | | Manual | N/A | N/A | N/A | N/A | -0.77 (0.24) | -0.015 | -3.19 | **0.001** | 0.46 (0.23) | 0.009 | 2.03 | **0.043** |

*coefficient. Time within Mode and Mode within Factor specified in a symbolic frame of Factor ⟹ Mode ⟹ Time. "N/A" stands for Manual Mode, and "Environment" is used as reference conditions.*

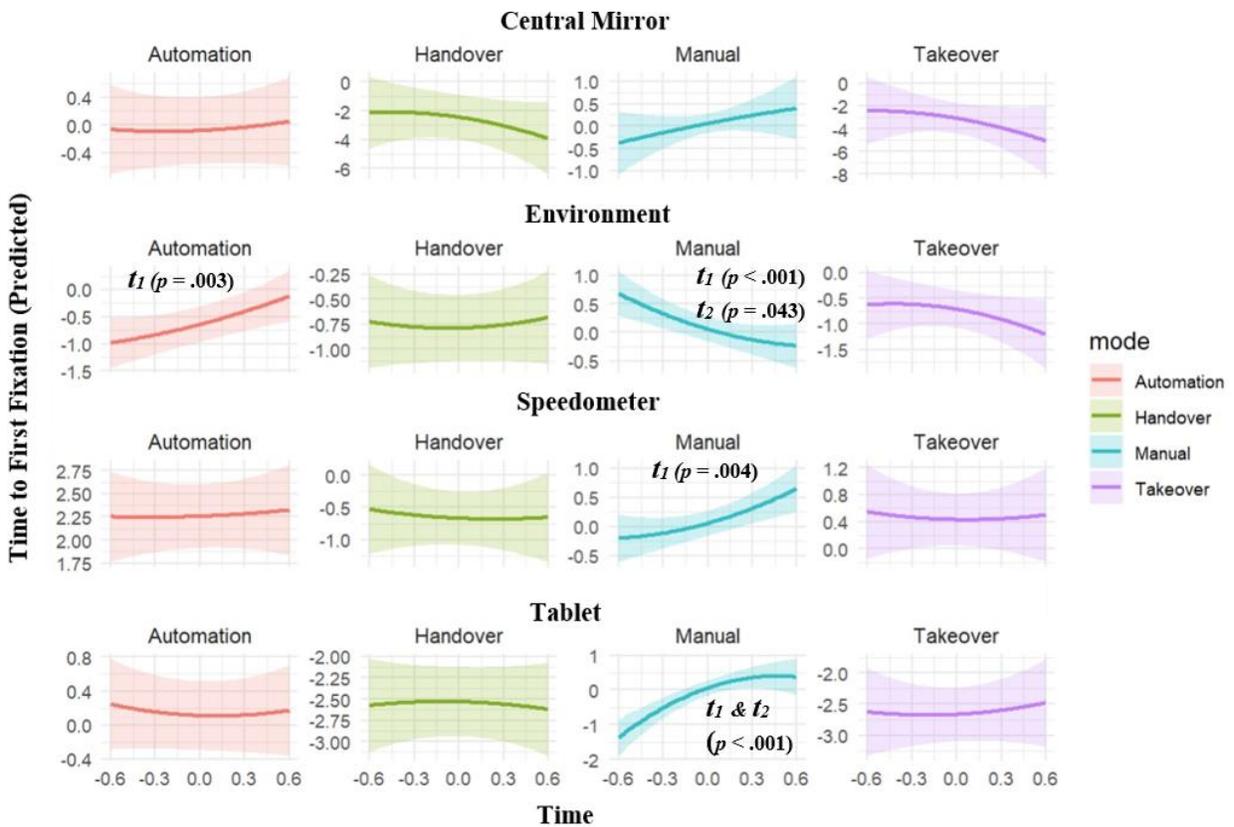

**Figure 5.** Predicted *TFF$_{Shift}$* over time across modes within factors. Time is represented on the x-axis as standardized and centered units (i.e., deviations from the mean). Shaded areas indicate %95 CI, highlighting variations in attention across modes within factors.

In *TFF$_{Shift}$*, following the significant fixed effects of mode within the Factor (Factor → Mode), pairwise contrast analyses were conducted to compare the different modes within each factor, revealing how quickly these areas attract attention, helping us understand the immediate visual priorities within the four factors, as summarized in

Table **7**. For the Environment factor, *TFF$_{Shift}$* was significantly higher in Manual mode compared to all the other three modes ($p < .001$), with no differences observed among the other three modes ($p = 1.000$). For the Environment factor, participants prioritized visual attention to the Environment more slowly in Manual mode, while the other modes showed equal and quicker attention allocation.

For the Tablet factor, *TFF_Shift* was significantly higher in Manual mode compared to Handover, Takeover, and Automation ($p < .001$). Automation mode also had significantly higher *TFF_Shift* compared to Handover and Takeover ($p < .001$), with no differences observed between Handover and Takeover ($p = 1.000$). The slower prioritization in Manual mode reflects reduced immediate attention to the Tablet, likely due to its lower relevance in manual control tasks, whereas quicker prioritization in Automation and transitional modes (Handover and Takeover) highlights its role in supporting task monitoring and transitions.

For the Speedometer factor, *TFF_Shift* was significantly lower in Manual mode compared to Automation ($p < .001$) and Handover ($p = .016$) but not significantly different from Takeover ($p = .667$). Automation mode had significantly higher *TFF_Shift* compared to Handover and Takeover ($p < .001$), and Handover was significantly lower than Takeover ($p < .001$). Faster prioritization in Manual and Handover modes indicates the critical importance of speed monitoring for immediate control tasks, while slower prioritization in Automation suggests reduced reliance on speed information under automated driving, with Takeover representing intermediate transitional priorities.

For the Central Mirror factor, *TFF_Shift* was significantly higher in Manual mode compared to Takeover ($p = .001$) but not significantly different from Automation or Handover ($p = 1.000$ and $p = .252$, respectively). Automation mode also had significantly higher *TFF_Shift* compared to Takeover ($p = .002$), with no significant differences observed between Handover and Takeover ($p = 1.000$). Slower prioritization in Manual mode suggests that participants relied less immediately on the mirror during active control, while faster prioritization in Handover and Takeover highlights the importance of immediate situational awareness during transitions, with Automation reflecting intermediate visual priorities.

Overall the mode-based effects for *TFF_Shift* reflect how participants' immediate visual priorities shift depending on the driving mode and factor. Slower prioritization in Manual mode across factors such as the Tablet and Central Mirror suggests that these areas were less critical during direct control, as participants focused on more immediate driving tasks like monitoring the Speedometer and Environment. In contrast, quicker prioritization in Handover and Takeover modes for factors like the Central Mirror and Tablet highlights the heightened need for SA and task transitions during these phases. Automation mode, characterized by slower prioritization of the Speedometer, indicates reduced reliance on certain visual zones due to the delegation of control to the system.

**Table 7**. Pairwise contrast analysis of $TFF_{shift}$ across modes nested within factors in the hierarchical structure.

| Factor | Contrast (Numerator / Denominator) | $\theta$ | $SE\ \theta$ | $\beta_{Std}$ | z-test | p-value |
|---|---|---|---|---|---|---|
| **Environment** | Manual / Automation | 2.11 | 0.27 | 0.11 | 5.85 | **< .001** |
| | Manual/ Handover | 2.39 | 0.31 | 0.12 | 6.73 | **< .001** |
| | Manual/ Takeover | 2.16 | 0.31 | 0.11 | 5.33 | **< .001** |
| | Automation / Handover | 1.13 | 0.15 | 0.06 | 0.97 | 1.000 |
| | Automation / Takeover | 1.02 | 0.15 | 0.05 | 0.16 | 1.000 |
| | Handover / Takeover | 0.90 | 0.13 | 0.05 | -0.72 | 1.000 |
| **Tablet** | Manual / Automation | 0.93 | 0.16 | 0.05 | -0.41 | 1.000 |
| | Manual/ Handover | 13.00 | 2.28 | 0.67 | 14.59 | **< .001** |
| | Manual/ Takeover | 15.01 | 2.78 | 0.77 | 14.64 | **< .001** |
| | Automation / Handover | 13.97 | 1.98 | 0.72 | 18.59 | **< .001** |
| | Automation / Takeover | 16.14 | 2.47 | 0.83 | 18.13 | **< .001** |
| | Handover / Takeover | 1.16 | 0.18 | 0.06 | 0.93 | 1.000 |
| **Speedometer** | Manual / Automation | 0.11 | 0.01 | 0.01 | -16.31 | **< .001** |
| | Manual/ Handover | 2.02 | 0.36 | 0.10 | 3.96 | **0.016** |
| | Manual/ Takeover | 0.67 | 0.11 | 0.03 | -2.51 | 0.667 |
| | Automation / Handover | 18.58 | 3.32 | 0.96 | 16.36 | **< .001** |
| | Automation / Takeover | 6.21 | 0.98 | 0.32 | 11.53 | **< .001** |
| | Handover / Takeover | 0.33 | 0.06 | 0.02 | -5.64 | **< .001** |
| **Central Mirror** | Manual / Automation | 1.13 | 0.25 | 0.06 | 0.58 | 1.000 |
| | Manual / Handover | 11.25 | 8.87 | 0.58 | 3.07 | 0.252 |
| | Manual / Takeover | 20.74 | 13.78 | 1.07 | 4.56 | **0.001** |
| | Automation / Handover | 9.91 | 7.70 | 0.51 | 2.95 | 0.326 |
| | Automation / Takeover | 18.27 | 11.89 | 0.94 | 4.46 | **0.002** |
| | Handover / Takeover | 1.84 | 1.84 | 0.10 | 0.61 | 1.000 |

**Note.** *"$\theta$" and "$SE\ \theta$" refer to the Ratio of two group means, "Mean Ratio" and its Standard Error, respectively, while "$\beta_{Std}$" denotes the standardized contrast coefficient. If $\theta > 1$, the numerator has a higher $TFF_{Shift}$ than the denominator.*

**Multivariate Regression Analysis for Predicting Dynamic Eye Behaviors by Trust**

We performed a multivariate regression analysis to investigate how trust in driving predicts eye-tracking behavior across various driving modes. The analysis was conducted using R software and its associated packages, following three key steps: *(1) Assessing skewness:* We evaluated the skewness of the three dependent eye-tracking variables using the e1071 package (Meyer et al., 2023). Variables with skewness values beyond ±1, indicating significant non-normality, were log-transformed to reduce skewness using the stats package (*Stats Package - RDocumentation*, n.d.) (Tabachnick & Fidell, 2006). After normalization, the skewness values fell within acceptable limits (±1), ensuring the data met the normality assumption required for regression analysis. (2) *Standardizing trust*: The independent variable, trust, was standardized using the dplyr (Wickham et al., 2014) and stats packages to achieve a mean of 0 and a standard deviation of 1. A quadratic term for trust (*tr²*) was also created to account for potential curvilinear effects, allowing us to explore how different levels of trust influence attention. The regression models included both the standardized trust variable and its quadratic term as predictors. (3) *Multivariate regression:* We conducted the multivariate regression using the broom (Robinson et al., 2014) and stats packages to predict each eye-tracking outcome based on the linear and quadratic terms of trust. The results include the standardized regression coefficient (*β*), its standard error (*SE β*), *t*-test values, *p*-values, and adjusted $R²$ ($R^2_{Adj}$) for both the linear and quadratic terms of trust.

The multivariate regression results revealed that trust had a significant effect on *fixation duration* in several driving modes, particularly in tasks involving the Tablet (see Table 8). In the automation mode, a significant positive quadratic relationship was found between trust and fixation duration, *β* = 0.18, *t*(2.33), *p* = .027, indicating that as trust increased, fixation duration also increased, with the effect becoming stronger at higher levels of trust. Similarly, in the handover mode, trust was found to be a significant predictor of fixation duration, *β* = 3.08, *t*(3.13), *p* = .004, where higher trust levels were associated with a substantial increase in fixation duration. Additionally, a significant quadratic effect in the handover mode, *β* = 2.68, *t*(3.49), *p* = .002, demonstrated that fixation duration increased more rapidly as trust reached higher levels. A similar pattern was observed in the takeover mode, where a significant quadratic relationship was found, $\beta = 2.03$, $t(2.36)$, $p = .026$, indicating that as trust increased, fixation duration also increased more rapidly at higher levels of trust. These findings suggest that trust plays a substantial role in determining fixation duration in tablet-related tasks, with higher trust leading to longer fixation times.

**Table 8.** Multivariate regression results predicting fixation duration by linear ($tr$) and quadratic ($tr^2$) trust terms.

| Factor | Mode ($R^2_{Adj}$) | Term | $\beta$ | SE | t | p |
|---|---|---|---|---|---|---|
| **Tablet** | Automation ($R^2_{Adj} = 0.184$) | tr | 0.02 | 0.10 | 0.20 | 0.843 |
| | | tr² | 0.18 | 0.08 | 2.33 | **0.027** |
| | Handover ($R^2_{Adj} = 0.344$) | tr | 3.08 | 0.99 | 3.13 | **0.004** |
| | | tr² | 2.68 | 0.77 | 3.49 | **0.002** |
| | Manual ($R^2_{Adj} = 0.032$) | tr | 0.08 | 0.09 | 0.96 | 0.348 |
| | | tr² | 0.03 | 0.07 | 0.45 | 0.659 |
| | Takeover ($R^2_{Adj} = 0.178$) | tr | 1.85 | 1.12 | 1.65 | 0.111 |
| | | tr² | 2.03 | 0.86 | 2.36 | **0.026** |
| **Environment** | Automation ($R^2_{Adj} = 0.071$) | tr | 4.80 | 3.23 | 1.49 | 0.148 |
| | | tr² | 1.83 | 2.57 | 0.71 | 0.482 |
| | Handover ($R^2_{Adj} = 0.022$) | tr | 2.70 | 3.47 | 0.78 | 0.442 |
| | | tr² | 0.53 | 2.76 | 0.19 | 0.849 |
| | Manual ($R^2_{Adj} = 0.049$) | tr | 4.40 | 3.60 | 1.22 | 0.231 |
| | | tr² | 1.82 | 2.86 | 0.64 | 0.530 |
| | Takeover ($R^2_{Adj} = 0.030$) | tr | 2.70 | 3.32 | 0.81 | 0.423 |
| | | tr² | -0.12 | 2.64 | -0.04 | 0.965 |
| **Speedometer** | Automation ($R^2_{Adj} = 0.049$) | tr | 0.29 | 0.34 | 0.87 | 0.393 |
| | | tr² | 0.31 | 0.27 | 1.15 | 0.262 |
| | Manual ($R^2_{Adj} = 0.003$) | tr | 0.32 | 1.14 | 0.28 | 0.782 |
| | | tr² | 0.15 | 0.90 | 0.16 | 0.870 |
| | Takeover ($R^2_{Adj} = 0.023$) | tr | 0.56 | 1.24 | 0.45 | 0.656 |
| | | tr² | -0.37 | 0.99 | -0.37 | 0.711 |
| **Central Mirror** | Automation ($R^2_{Adj} = 0.071$) | tr | -0.09 | 0.07 | -1.22 | 0.234 |
| | | tr² | -0.03 | 0.06 | -0.47 | 0.644 |

Note. "$\beta$" represents the standardized regression coefficient, and "SE $\beta$" refers to its standard error.

Trust also had a significant effect on fixation count, particularly in the Tablet, see Table 9. In the automation mode, a significant quadratic relationship was found between trust and fixation count, $\beta = 0.01$, $t(2.24)$, $p = .033$, indicating that as trust increased, participants fixated on the Tablet more frequently, with the effect becoming more pronounced at higher levels of trust. Similarly, a significant quadratic effect was observed in the handover mode, $\beta = 0.04$, $t(2.25)$, $p = .033$, showing that fixation count increased as trust levels increased, particularly at higher trust levels. These results suggest that as trust increases, participants engage more frequently with tablet-related AOIs, particularly in automation and handover modes, with stronger effects observed at higher trust levels.

**Table 9.** Multivariate regression results predicting fixation count by linear ($tr$) and quadratic ($tr^2$) trust terms.

| Factor | Mode ($R^2_{Adj}$) | Term | $\beta$ | $SE$ | $t$ | $p$ |
|---|---|---|---|---|---|---|
| **Tablet** | Automation ($R^2_{Adj} = 0.162$) | tr | 0.00 | 0.00 | 0.40 | 0.689 |
| | | $tr^2$ | 0.01 | 0.00 | 2.24 | **0.033** |
| | Handover ($R^2_{Adj} = 0.153$) | tr | 0.02 | 0.02 | 1.14 | 0.262 |
| | | $tr^2$ | 0.04 | 0.02 | 2.25 | **0.033** |
| | Manual ($R^2_{Adj} = 0.037$) | tr | 0.01 | 0.01 | 1.03 | 0.311 |
| | | $tr^2$ | 0.00 | 0.01 | 0.51 | 0.613 |
| | Takeover ($R^2_{Adj} = 0.050$) | tr | 0.01 | 0.02 | 0.31 | 0.759 |
| | | $tr^2$ | 0.02 | 0.01 | 1.15 | 0.259 |
| **Environment** | Automation ($R^2_{Adj} = 0.001$) | tr | 0.00 | 0.02 | 0.15 | 0.878 |
| | | $tr^2$ | 0.00 | 0.01 | 0.13 | 0.900 |
| | Handover ($R^2_{Adj} = 0.009$) | tr | -0.01 | 0.04 | -0.32 | 0.749 |
| | | $tr^2$ | -0.01 | 0.03 | -0.50 | 0.624 |
| | Manual ($R^2_{Adj} = 0.010$) | tr | -0.01 | 0.02 | -0.42 | 0.679 |
| | | $tr^2$ | 0.00 | 0.02 | 0.09 | 0.932 |
| | Takeover ($R^2_{Adj} = 0.023$) | tr | 0.02 | 0.04 | 0.43 | 0.673 |
| | | $tr^2$ | -0.01 | 0.03 | -0.41 | 0.685 |
| **Speedometer** | Automation ($R^2_{Adj} = 0.031$) | tr | 0.00 | 0.01 | 0.56 | 0.583 |
| | | $tr^2$ | 0.00 | 0.00 | 0.94 | 0.357 |
| | Handover ($R^2_{Adj} = 0.129$) | tr | 0.10 | 0.06 | 1.74 | 0.095 |
| | | $tr^2$ | 0.06 | 0.06 | 0.97 | 0.342 |
| | Manual ($R^2_{Adj} = 0.023$) | tr | 0.01 | 0.02 | 0.80 | 0.431 |
| | | $tr^2$ | 0.01 | 0.01 | 0.58 | 0.567 |
| | Takeover ($R^2_{Adj} = 0.003$) | tr | 0.00 | 0.01 | -0.19 | 0.851 |
| | | $tr^2$ | 0.00 | 0.01 | -0.25 | 0.801 |
| **Central Mirror** | Automation ($R^2_{Adj} = 0.024$) | tr | 0.00 | 0.00 | -0.71 | 0.482 |
| | | $tr^2$ | 0.00 | 0.00 | 0.01 | 0.995 |

Note. "$\beta$" represents the standardized regression coefficient, and "SE $\beta$" refers to its standard error.

The relationship between trust and time to first fixation was also significant in both the Tablet and central mirror, summarized in Table 10. In the automation mode, a significant negative quadratic relationship was found between trust and time to first fixation on the Tablet, $\beta$ = -0.22, $t(-3.45)$, $p$ = .002. As trust increased, participants took less time to fixate on the Tablet, and this effect became more pronounced at higher trust levels, indicating that higher trust led to quicker visual engagement with the Tablet. Conversely, in the central mirror, a significant positive relationship was found between trust and time to first fixation in the automation mode, $\beta$ = 0.24, $t(2.41)$, $p$ = .023. Higher trust levels were associated with longer times to first fixation on the central mirror, suggesting that as trust increased, participants reduced their attention to monitoring like the central mirror.

**Table 10.** Multivariate regression results predicting time to first fixation by linear (tr) and quadratic (tr$^2$) trust terms.

| Factor | Mode ($R^2_{Adj}$) | Term | $\beta$ | $SE$ | $t$ | $p$ |
|---|---|---|---|---|---|---|
| **Tablet** | Automation ($R^2_{Adj} = 0.341$) | tr | -0.01 | 0.08 | -0.11 | 0.910 |
| | | tr$^2$ | -0.22 | 0.06 | -3.45 | **0.002** |
| | Handover ($R^2_{Adj} = 0.044$) | tr | -0.12 | 0.10 | -1.12 | 0.272 |
| | | tr$^2$ | -0.06 | 0.08 | -0.75 | 0.462 |
| | Manual ($R^2_{Adj} = 0.042$) | tr | -1.49 | 3.52 | -0.42 | 0.675 |
| | | tr$^2$ | 1.89 | 2.75 | 0.69 | 0.497 |
| | Takeover ($R^2_{Adj} = 0.047$) | tr | -0.02 | 0.10 | -0.20 | 0.846 |
| | | tr$^2$ | -0.08 | 0.08 | -1.08 | 0.291 |
| **Environment** | Automation ($R^2_{Adj} = 0.033$) | tr | -0.01 | 0.10 | -0.11 | 0.916 |
| | | tr$^2$ | -0.07 | 0.08 | -0.92 | 0.367 |
| | Handover ($R^2_{Adj} = 0.024$) | tr | -0.07 | 0.09 | -0.75 | 0.459 |
| | | tr$^2$ | -0.05 | 0.07 | -0.71 | 0.481 |
| | Manual ($R^2_{Adj} = 0.011$) | tr | -0.03 | 0.12 | -0.22 | 0.831 |
| | | tr$^2$ | -0.05 | 0.10 | -0.57 | 0.570 |
| | Takeover ($R^2_{Adj} = 0.148$) | tr | -0.09 | 0.06 | -1.62 | 0.116 |
| | | tr$^2$ | 0.03 | 0.05 | 0.59 | 0.563 |
| **Speedometer** | Automation ($R^2_{Adj} = 0.026$) | tr | -1.65 | 3.14 | -0.52 | 0.604 |
| | | tr$^2$ | -2.13 | 2.50 | -0.85 | 0.401 |
| | Handover ($R^2_{Adj} = 0.017$) | tr | 0.02 | 0.12 | 0.14 | 0.893 |
| | | tr$^2$ | -0.07 | 0.13 | -0.53 | 0.602 |
| | Manual ($R^2_{Adj} = 0.033$) | tr | -0.12 | 0.17 | -0.71 | 0.483 |
| | | tr$^2$ | -0.13 | 0.13 | -0.95 | 0.352 |
| | Takeover ($R^2_{Adj} = 0.013$) | tr | 0.01 | 0.09 | 0.11 | 0.914 |
| | | tr$^2$ | -0.03 | 0.07 | -0.48 | 0.638 |
| **Central Mirror** | Automation ($R^2_{Adj} = 0.179$) | tr | 0.24 | 0.10 | 2.41 | **0.023** |
| | | tr$^2$ | 0.07 | 0.08 | 0.86 | 0.398 |

| | | | | | |
|---|---|---|---|---|---|
| Manual | tr | -5.14 | 5.78 | -0.89 | 0.386 |
| ($R^2_{Adj}$ = 0.051) | tr$^2$ | -4.41 | 5.46 | -0.81 | 0.430 |

Note. "$\beta$" and "$SE\ \beta$" refer to the standardized regression coefficient and its Standard Error, respectively.

## DISCUSSION AND CONCLUSION

This study addressed two primary research questions to enhance our understanding of the cognitive demands of different driving modes and their implications for designing human-machine interfaces in automated vehicles. The first question examined how attention, as measured by eye-tracking, varies across different AOIs related to both technological and environmental factors during four distinct driving modes: *automation, Handover, manual, and takeover*. The findings support the first hypothesis that attention varies significantly across AOIs depending on the driving mode. For instance, in Manual mode, participants consistently allocated greater attention to environmental factors, which aligns with the hypothesis that the higher cognitive demands of active driving require sustained environmental focus, but the attention shifts were not as dynamic as expected. This is consistent with Pipkorn et al. (2024)'s study that repeated exposure to takeover requests (TORs) did not affect environmental attention during manual driving. In Automation mode, on the other hand, fixation durations on technological factors, such as the Tablet, remained stable, as hypothesized, reflecting the lower cognitive demands and consistent reliance on automated systems. Further supporting this finding, Guo et al. (2024)'s study indicated that attention on the Tablet was primarily tied to assessing the necessity of a takeover. During Handover mode, as expected, participants dynamically shifted their attention between technological and environmental factors, reflecting the gradual transition of control from vehicle to the driver. Takeover mode, as hypothesized, exhibited the most dynamic attention shifts, with participants rapidly reallocating their focus between AOIs to regain control of the vehicle. Merat et al. (2014) also found a great degree of variability in attention during taking over, especially when the transition was unexpected or triggered by distractions. This variability in attention led to slower stabilization of visual focus and driving control, highlighting the challenges drivers face in regaining situational awareness when transitioning abruptly from automated to manual control.

The second question examined how trust in automation influences attention allocation to environmental and technological factors across the four driving modes. Trust significantly influenced fixation duration, fixation count, and time to first fixation, especially in tablet-related AOIs. As hypothesized and consistent with previous studies (Demir et al., 2021; N. Tenhundfeld et al., 2022; N. L. Tenhundfeld et al., 2019), the quadratic effects of trust reveal the nonlinear nature of this relationship, where higher levels of trust lead to more pronounced effects. The results confirmed that as trust in automation increased, participants allocated less attention to monitoring-related AOIs, such as the speedometer and central mirror, across automation, Handover, and takeover modes. Increased trust, on the other hand, led to quicker engagement with the Tablet. These findings suggest that with greater trust, participants' attention shifts from monitoring-related tasks to tablet-related or secondary tasks. This pattern aligns with findings from the previous study by Hergeth et al. (2016), who similarly observed that participants with higher levels of trust in automation monitored the system less frequently. In handover and takeover modes, higher trust was associated with a faster time to first fixation on secondary tasks, suggesting increased confidence in managing vehicle control transitions.

**Eye-Tracking Measures**

Fixation Duration measures the total time a participant's gaze remains fixed on specific AOIs, while Fixation Count tracks how often the gaze shifts to these areas. The result shows that Fixation duration and fixation count has similar pattern as they both provide insights into how participants distribute and allocate their visual attention across different AOIs, which is consistent with previous study by Liang et al. (2021)

For the environmental factors, fixation duration showed significant time-related changes in manual mode, indicating increased attention over time. This increasing attention may stem from the need to maintain situational awareness, as suggested by the Attention-Situation Awareness Model (Bellenkes et al., 1997), which posits that situation awareness guide attention allocation. High levels of attention can help maintain strong situational awareness. In contrast, drivers in automation mode tended to monitor the road less over time, allocating reduced attention to the driving environment. In takeover mode, fluctuations and increases in attention to environmental factors were observed over time, consistent with findings by Gold et al. (2013) that drivers needed additional time to regain situational awareness when transitioning from automation, leading to improved focus on the driving environment.

For the speedometer factor, a similar pattern was observed. In takeover mode, both fixation duration and fixation count showed significant increases over time. This suggests that as participants regained control, they focused more on the speedometer. It could be interpreted that drivers had to manually control the speed of vehicle, with a greater attention allocation on speedometer. The result of Pipkorn et al. (2024) also reported a significant increasing of the percentage of fixation duration on the speedometer in take over mode. This suggests that during the taking over process, drivers momentarily prioritized speed monitoring over environmental awareness, which could temporarily delay their focus on the road.

***Time to First Fixation.***

Time to First Fixation is the time elapses from the beginning of a visual stimulus or scene until the participant's gaze fixates on a specific AOI. It reflects how quickly these areas attract attention, helping us understand the immediate visual priorities within each of the four factors.

For the environmental factors, no significant differences were found across modes, indicating consistent attention. For the Tablet, participants in Automation mode showed quicker fixation times over time, likely due to reduced cognitive demand of driving, while Manual mode had more longer times to first fixation, suggesting increased cognitive load that driver concentrated on the road and do not have time to focus on the Tablet. This implies that automation made visual processing easier, while manual tasks required more sustained attention (Y. Ma et al., 2020) illustrates uncompetitive relationships between visual attention and cognitive attention. It means that drivers need to take mental resources to interpret the visual information related to driving task in manual mode. As the degree of cognitive workload increases, cognitive attention occupies resources more than the visual.

For the speedometer factor, significant fluctuations were observed only in Automation mode, indicating varying focus with task complexity, while other modes showed stable attention, suggesting lower visual demands. For the central mirror factor, both Automation and Manual modes exhibited fluctuating times to first fixation, with Manual mode showing a sharp decrease later, possibly due to task fatigue or divided attention. Handover and Takeover modes remained stable, reflecting lower cognitive demands in these conditions. Winter et al. (2016) found that during higher levels of unreliable automation, participants experienced increased frustration, which influenced their attention by causing fluctuating focus and heightened vigilance.

**Implications**

At Level 3 or higher of automated driving, drivers are no longer required to monitor the vehicle or the road, allowing them to participate in NDRTs (Kim et al., 2021). However, drivers still need to be vigilant to take over control of the vehicle in occasional cases where the automated system fails (Sever & Contissa, 2024). We conducted our simulation in the absence of NDRTs, and participants needed to respond to a TOR when necessary. The results of attention allocated to the environment factor and the tablet factor play an important role in explaining drivers' cognitive load. Daniel Kahneman (Kahneman, 1973) proposed the theory of capacity-limited cognitive resources which is that individuals must expend their cognitive resources to process information about external events. In our findings, drivers tend to maintain stable attention to the tablet factor during autonomous driving, with fluctuations occurring when a TOR was issued. We suppose that drivers regard the Tablet as a stable factor during autonomous driving while drivers regard the environment as a dynamic factor. Thus, the driver's cognitive resources were allocated at a stable level to the Tablet but shifted dynamically when a TOR was required. Greater attention was allocated to the environment during manual driving, which represents a high cognitive load required to process the surrounding information. Wintersberger et al. (2018) found that issuing TORs at natural task boundaries improves takeover performance by reducing cognitive load. We could imply that autonomous systems should deliver TORs at optimal times based on drivers' engagement levels to ensure smoother transitions, minimize erratic attention shifts, and enhance overall safety. TORs had a significant impact on the dynamics of drivers' attention, as they increased driver workload. Multiple TORs improved driver acceptance of operations on the Tablet and caused decreased attention to the Tablet during autonomous driving, which is consistent with previous study by Hecht et al. (2020). However, Hecht et al. did not find a significant effect on trust after frequent TORs, interpreting this as evidence that trust may not be as sensitive to takeover actions compared to the system's overall reliability or participants' familiarity with the automated system. Furthermore, Merat et al (2014) found that drivers' visual attention was more erratic when automation disengaged due to distractions compared to predictable, system-based disengagement. This variability can be linked to the fluctuations in fixation count during certain modes in our study, indicating how unpredictability in the control environment affects drivers' ability to maintain consistent visual focus.

### *Practical*

Adaptive HMI were investigated by previous study by Ulahannan et al. (2020).With drivers' attention patterns well described, the HMI design are able to take advantage of dynamics of attention and present the most appropriate information. Different ratios of drivers' attention on different AOI reveal the potential to optimize drivers' monitoring task through well-designed HMI layout. It can be achieved by prioritizing information based on its importance and relevance to the driving context. For example, during high demand driving situations, the interface can adapt to show only the most critical information, reducing distraction, cognitive load and enhancing usability (Korthauer et al., 2020). The result of our study suggests that drivers allocated less attention to the environment but greater attention to tablets. It means that more information and explanations about what the vehicle was doing and why it was doing it, should be added to HMI tablet. The traffic situation behind the vehicle could be added to HMI tablet, as drivers tend to allocate less attention to the central mirror.

### Limitations of the Study

During long periods of manual driving, drivers primarily focus on monitoring road conditions while utilizing their spare attentional capacity for processing other information is noteworthy (Hughes & Cole, 1986). This suggests that drivers may allocate visual attention differently based on task demands and available cognitive resources (Birrell & Fowkes, 2014) . However, our study did not explicitly consider the factor of human spare vision, which may introduce errors in the analysis of gaze behavior during manual driving.

As for level 3 autonomous vehicles, drivers are still required to monitor road conditions although they don't have to manually control the vehicle for most of the time. Prolonged supervision in this state can lead to boredom, monotony, and fatigue, potentially impacting attention, and safety (Coughlin et al., 2011). Our experiment, which periodically transitioned drivers into autonomous driving states for several minutes, did not capture the changes in visual behavior resulting from long-term monitoring. This omission may limit the generalizability of our findings in L3 autonomous driving scenarios.

## CONCLUSION

The study explored the cognitive demands and attention dynamics across different driving modes in automated vehicles, focusing on how drivers allocate their attention and how trust in automation influences this allocation. Key findings include:

- **Attention Allocation**: Drivers allocate attention differently based on driving modes. In manual mode, drivers focus more on environmental factors, whereas in automation mode, attention shifts to technological factors (e.g., tablets). During Handover and takeover, attention shifts dynamically, highlighting the challenges drivers face when regaining control.

- **Trust and Attention**: Higher trust in automation resulted in reduced attention to monitoring tasks and a greater focus on secondary tasks, such as tablet interactions. Trust in automation influences how quickly drivers respond during control transitions.

FUTURE DIRECTIONS

The integration of physiological measures (e.g., heart rate variability, electroencephalogram) will be explored as complementary metrics to assess the cognitive workload of drivers. These measures can provide a more comprehensive understanding of the physiological and psychological state of drivers during automated driving, allowing researchers to assess cognitive workload in real-time and develop better adaptive systems.

Furthermore, attention is sometimes interpreted as a cognitive load about information processing. When cognitive resources are limited, arousal level determines the amount of available cognitive resources. Arousal levels vary between individuals and are influenced by cognitive capacity, the higher the arousal level, the greater the amount of available cognitive resources. Arousal, which refers to a driver's responsiveness to stimuli, can be investigated to examine its relationship with takeover performance in autonomous vehicles. Understanding how arousal levels affect drivers' ability to regain control of the vehicle during a takeover event can help in designing intervention strategies that optimize drivers' readiness and improve takeover outcomes. This could help design adaptive alerts systems that help drivers maintain an optimal arousal level, ensuring a smoother transition from automation to manual control.

Finally, longitudinal studies that track how driver trust and attention patterns evolve over time with increased exposure to automated systems would be valuable. Such research could reveal how familiarity with automated driving influences drivers' cognitive workload, trust levels, and attention allocation, ultimately informing the design of systems that better support driver engagement and safety over time.

## ACKNOWLEDGEMENTS

TBD: provide if there is any